\documentclass[%
 reprint,
 amsmath,amssymb,
 aps,
pra,
showkeys,
]{revtex4-2}
\usepackage{graphicx}
\usepackage{subfig}
\graphicspath{ {./imgs/} }
\usepackage{bm}
\usepackage{subfig}
\usepackage{xcolor}
\usepackage{cases}
\usepackage{appendix}

\newcount\myloopcounter
\newcommand{\repeatit}[2][10]{
  \myloopcounter0
  \loop\ifnum\myloopcounter < #1
  #2
  \advance\myloopcounter by 1
  \repeat
}

\begin{document}
\title{Collective tunnel ionization in atomic systems}

\author{D.I. Tyurin}
    \email[Correspondence email address: ]{denisturin1999@yandex.ru}
    \affiliation{National Research Nuclear University MEPhI, Kashirskoe shosse 31, 115409, Moscow, Russia}

\author{V.V. Strelkov}
    \affiliation{ P.N. Lebedev Physical Institute of the Russian Academy of Sciences, Leninsky prospect 53, 119991, Moscow, Russia}

\author{S.V. Popruzhenko}
    \affiliation{National Research Nuclear University MEPhI, Kashirskoe shosse 31, 115409, Moscow, Russia}
    \affiliation{Prokhorov General Physics Institute of the Russian Academy of Sciences, Vavilova str. 38, 119991, Moscow, Russia}


\date{\today} 

\begin{abstract}
We present a theory of the collective tunnel effect in atomic systems subject to strong low-frequency laser fields.
Using an analytic semi-classical method and numerical solution to the time-dependent Schr\"{o}dinger equation we demonstrate the collective channel of the nonsequential double ionization in a model two-dimensional two-electron system with ionization potentials reproducing those of neutral xenon and of the negative bromine ion.
Collective tunneling in the presence of the electron-electron repulsion comprises the joint sub-barrier motion of the two electrons along quantum trajectories transversely shifted in opposite directions with respect to the direction of the external electric field.
Analytic results for the rates of collective tunneling allow indicating the intervals of laser intensity, where this effect can be expected to compete with sequential tunneling in helium, xenon, and negative bromine.
Momentum distributions of the electron pairs carry clear signatures of the collective channel in the form of shoulders in the direction lateral to the field polarization.

\end{abstract}


\maketitle

\section{Introduction}
Nonsequential double ionization (NSDI) of atoms and molecules is one of the remarkable manifestations of the highly nonlinear, nonperturbative nature of light-matter interactions at high laser intensities.
The pioneering laboratory observation made in the 70-es \cite{suran_jtp75} for atoms of alkali metals, was followed by a number of high-quality experiments \cite{huillier_pra83,kulander_prl92,walker_prl94,kondo_pra93,cornaggia_jpb98} studied NSDI in noble gases and molecules.
In strong low-frequency laser fields, where single and multiple ionization is possible through simultaneous absorption of a large number of laser photons, the common pathway of multiple ionization is sequential detachment of electrons from their atomic residual.
Intuitively, one expects this sequential ionization (SI) channel  to dominate the ionization dynamics.
However, the pivotal experiments (see e.g. \cite{huillier_pra83,kulander_prl92,walker_prl94,weber_prl00,moshammer_prl00,rudenko_jpb08} and the review \cite{becker-rmp12} for further references), as well as the bulk of studies performed later shown that, in linearly polarized fields, the yield of doubly charged ions may exceed that predicted within the picture of SI by several orders of magnitude.
The same phenomenon was observed for multiple ionization, when three or more electrons are being removed from the atom \cite{chin_jpb98}.

The dominance of nonsequential ionization mechanisms in the production of multiply charged ions at laser intensities $10^{14}-10^{17}{\rm W/cm}^2$, wavelengths $\lambda\simeq 1\mu{\rm m}$ and polarization close to linear is by now out of doubt.
A particular scenario, which can be responsible for this giant nonlinear effect, remained under intense debates before the 2000-es, when the measurement of momentum distributions of doubly and triply charged ions and of photoelectron pairs \cite{weber_prl00,moshammer_prl00} gave  decisive arguments in favor of the recollision mechanism \cite{kuchiev-jetpl87,corkum-prl93}. 
In simple words, the latter assumes that the outermost (i.e., the one having the lowest ionization potential $I_p$) electron removed from the atom though multiphoton or tunnel ionization can be driven back to its parent ion and collisionally ionize or excite one or several bound electrons.
Calculations performed within several approximate quantum-mechanical models built on the basis of the recollision concept, have shown the efficiency of this channel, explained its great dominance over the sequential one, and reproduced qualitatively the experimentally recorded momentum distributions \cite{kopold_prl00,becker-rmp12}.
By now, the recollision scenario of NSDI has been thoroughly studied both experimentally and theoretically; see Ref.~\cite{becker-rmp12} for a review.
Except for the NSDI, the recollision mechanism underlies such effects as the generation
of high-order harmonics (HHG) and high-energy plateau of above-threshold ionization. 
The former is known to be the base stone of the currently thriving area of attosecond science \cite{agostini_nobel24,huillier_nobel24,krausz_nobel24}.

Since then, other hypothetic nonsequential ionization scenarios, including shake-off \cite{agostini-rev95} and collective tunneling \cite{zon-jetp99,becker-prl00} remained obscured by recollision, both because of challenges in experimental verification and the absence of an appealing and consistent theoretical treatment.
The collective tunneling models~\cite{zon-jetp99,becker-prl00,tyurin-leb23} have been theoretically and numerically examined \cite{lein_prl00, becker-prl00} without coming to certain quantitative conclusions.
No signatures of these hypothetical channels have been experimentally evidenced so far. 
The primary obstacle is the prevailing dominance of the recollision channel in linearly polarized fields. 
Therefore, to uncover alternative mechanisms of correlated ionization, one should find a way to suppress recollisions.
This can be achieved, for instance, by using circularly or elliptically polarized fields, where the two‑dimensional motion of the electrons prevents its return to the parent ion \cite{keitel-pra07, kolya-pra08}.
Another way is to employ extremely short laser pulses with duration insufficient to support recollisions \cite{carla_prl04,paulus_jpb06, bergues_nc12,kubel_pra13,ben_oe16,paulus_jpb18,itzhak_nc20}.
While such pulses are challenging to produce experimentally, they are readily accessible in numerical modeling, offering an additional tool for uncovering weak correlation effects that would otherwise be hidden under the consequences of recollision dynamics.

In this paper, we demonstrate through numerical solutions to the time-dependent Schr\"{o}dinger equation (TDSE) that, in extremely short pulses supporting no recollision, a contribution of the collective channel can be identified in the flux of the two-electron wave function.
This collective flux associated with a simultaneous escape of two electrons, appears a two-dimensional (2D) phenomenon, which becomes possible only due to the lateral, with respect to the ionizing electric field, motion of the electron pair.

Our central task is to build an analog of the celebrated theory of single-electron tunnel ionization \cite{opp_pr28,keldysh-jetp65,popov_jetp66}, but for the case when two electrons tunnel out of the atom at the same time.
Quite surprisingly, this problem appears, to a great extent, analytically resolvable despite the electron-electron repulsion, which plays the key role and can not be discarded.
In other terms, we present a particular solution of the reduced three-body quantum problem in the case, when two light interacting particles tunnel from one part of the classically-allowed space to another. 
It is remarkable that the very concept of two-particle tunneling has been pioneered long ago in the theory of two-proton radioactivity of nuclei \cite{zeldovich_jetp60,galitsky_np64}, see also review \cite{blank_rpp08} for the state-of-the art.
In this way, our study bridges two conceptually similar phenomena in atomic and nuclear physics.

To investigate the phenomenon of collective tunneling (CT), we solve numerically the two‑electron 2D (for each electron) TDSE for model systems with ionization potentials of neutral xenon and of the negative bromine ion and show that the effect is present in both systems.
This demonstration is important in view of a possible experimental search, for noble gases are known to be suitable targets in strong field ionization experiments. 
We develop a semiclassical model of CT, which, in contrast to the earlier theories \cite{zon-jetp99,becker-prl00}, takes the electron-electron interaction into account.
This account requires a considerable extension of the Imaginary Time Method \cite{popov-usp04,popov2005} or the formalism of quantum orbits \cite{salieres_sci01} customary in the theory of strong-field atomic and molecular phenomena.
We demonstrate that our analytic model captures the key numerically observed features of the coordinate and momentum distributions of the electron pairs released through simultaneous tunneling. 

The paper is organized as follows.
In the next section, we sketch the idea and use qualitative arguments to convince the reader that the collective channel of tunnel ionization can effectively contribute to the rate of double ionization in a strong low-frequency laser field.
In Section III, we formulate the statement of the problem and describe the numerical approach for solving the two-electron Schr\"{o}dinger equation in 1D and 2D. 
Section IV reports numerical results for neutral xenon and the negative bromine ion.
In Section V we present our analytic semiclassical theory of collective tunneling.
In the next Section we consider the CT rates and lateral momentum distributions of photoelectron pairs, looking for possible experimentally detectable signatures of collective tunneling.
The last Section contains conclusions and outlook.
Technical details of numerical and analytic calculations are described in Appendices.
Atomic units $\hbar=m_e=|e|$ are used.
Throughout the text, we use acronyms SI for sequential ionization and CT for collective tunneling.

\section{Collective tunnel ionization in a nutshell}

Before addressing details of numerical solutions, which demonstrate the collective electron flux induced by a strong external field and of an analytic theory describing this effect, we would like to present some simple considerations, which suggest that such a process as a simultaneous penetration of two electrons through the potential barrier created by their interaction with the ion, the electric field and with each other can take place with a probability comparable to that of single-electron tunneling.
Consider a two-electron atom or ion, the electron coordinates are $\textbf{r}_1$ and $\textbf{r}_2$, the electron-electron repulsion is $V_{\rm ee}(\textbf{r}_1,\textbf{r}_2)$, the interaction of each electron with the parent ion is $V_{\rm ei}(\textbf{r}_{1,2})$, and the interaction of each electron with the external electric field is $-{\bf E}\cdot{\bf r}_{1,2}$. 
Double ionization in a static or slow-varying field ${\bf E}$ is defined by the tunneling of the electrons through the barrier 
\begin{equation} 
V(\textbf{r}_{1}, \textbf{r}_{2})= V_{\rm ei}(\textbf{r}_{1})+V_{\rm ei}(\textbf{r}_{2})+V_{\rm ee}(\textbf{r}_1,\textbf{r}_2) -{\bf E}\cdot{\bf r}_1 -{\bf E}\cdot{\bf r}_2~.
\label{V_general}
\end{equation}
The interactions in (\ref{V_general}) can be smoothed at small distances to adjust the ionization potentials of the model two-electron system to those of a real atom or ion (see details in Section III below), while for large distances they behave as the Coulomb attraction to a nucleus with charge $Z$ and the electron-electron Coulomb repulsion, correspondingly:
\begin{equation}
    V_{\rm ei}({\bf r}_{1,2})=-\frac{Z}{r_{1,2}}~,~~~
    V_{\rm ee}({\bf r}_1,{\bf r}_2)=\frac{1}{|{\bf r}_2-{\bf r}_1|}~.
\label{V_asympt}
\end{equation}

As we are interested in the {\it collective}, simultaneous escape of the electrons, it would be reasonable to assume (we will prove this assumption in Sections IV and V below) that they move having close projections of their coordinates on the field direction, which we denote as $X$, $x_1\approx x_2$. 
The relative coordinate ${\bf r}_{12}={\bf r}_2-{\bf r}_1$ will then be approximately orthogonal to the field ${\bf E}$, while the symmetry arguments suggest that the lateral with respect to the field coordinates will rather be opposite, ${\bf r}_{\bot 1}\approx -{\bf r}_{\bot 2}$.
The system of two electrons in the atom has the total energy $-I_p=-(I_{p1}+I_{p2})$, where $I_{p1}$, and $I_{p2}$ are the first and the second ionization potentials, respectively. 
Adopting this naive picture where the center of mass moves along the electric field with coordinate $X$, and ${\bf r}_{12}\bot{\bf E}$ so that ${\bf r}_{\bot 2}=-{\bf r}_{\bot 1}={\bf r}_{12}/2\equiv {\bf r}_{\bot}$, we can study the potential energy (\ref{V_general}) on this surface as function of $X$ and $r_{\bot}$.
The tunneling probability is defined by the barrier height and width of the classically forbidden domain where
\begin{equation} 
V(X,r_{\bot})>-I_p
\label{collective_condition_general}
\end{equation}
Taking into account Eq.~(\ref{V_general}) the condition~(\ref{collective_condition_general}) is written as:
\begin{equation}
 V_{\rm ei}\bigg(\sqrt{X^2+r_{\bot}^2}\bigg) -EX>-\frac{1}{2}I_p - \frac{1}{2}V_{\rm ee}(r_{\bot})
 \label{collective_condition}
\end{equation}

In the sequential ionization picture, the double ionization probability is mostly defined by that for the inner electron, assuming the outer one is already detached. 
For this process, the sub-barrier region is determined by the condition:
\begin{equation}
 V_{\rm ei}\bigg(\sqrt{x^2_2+r_{\bot 2}^2}\bigg) -Ex_2>-I_{p2},   
  \label{sequential_condition}
\end{equation}
here ${\bf r}_2=(x_2,{\bf r}_{\bot 2})$ is the inner electron coordinate. 
In this tutorial case, the barrier height and width are the smallest when $r_{\bot 2}=0$, while for collective tunneling this is not the case, because $V_{\rm ee}$ grows for small $r_{\bot}$.
This fundamental difference between the topology of single-electron and collective tunneling is illustrated in Fig.1, where we plot the potential (\ref{V_general}) on the surface $(Y=0,{\bf r}\cdot E=0)$ as a function of $X$ and $r_{\bot}$ and the corresponding single-electron potential for the inner electron.
Note that a similar analysis of the simultaneous two-electron escape has been performed in Refs.\cite{eckhardt_epj01,eckhardt_pra01,eckhardt_jpb06} by Eckhardt and Sacha within the classical paradigm of the over-the-barrier ionization.

\begin{figure}[!htb]
\centering
\includegraphics[width=0.9\linewidth]{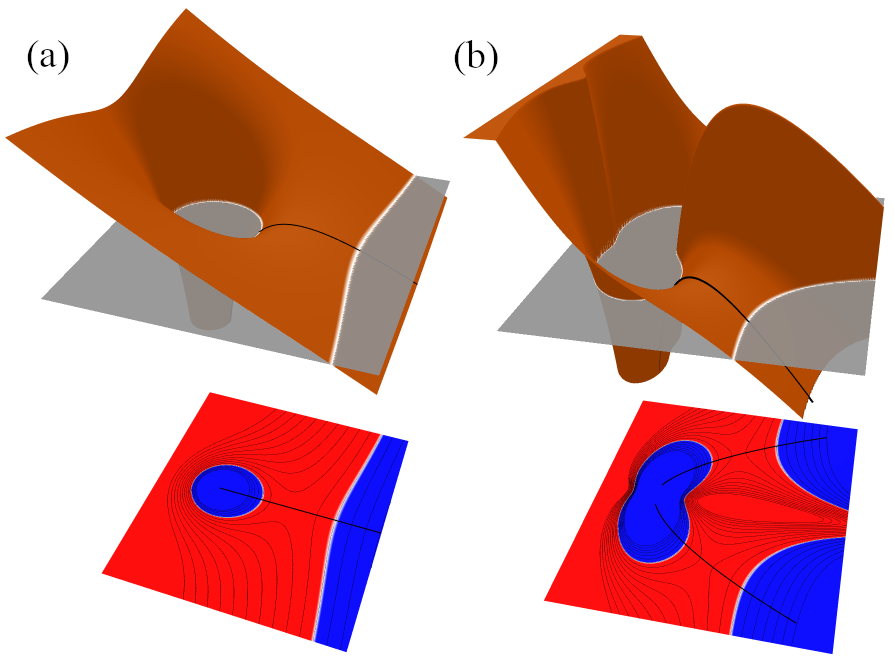}
\caption{Single- and two-electron potential energies (upper row) and the corresponding classically forbidden (red) and allowed (blue) parts of space (bottom row) plotted for the two first ionization potentials of Xe, $I_{p1}=12.1$eV, $I_{p2}=21.2$eV and the static electric field $E=0.06$ directed along the $x$-axis. Column (a): the single-electron potential in coordinates $x_2,r_{\bot 2}$ for $I_p=I_{p2}$; column (b): the two-electron potential (\ref{V_general}) with $I_p=I_{p1}+I_{p2}$, $Z=2$ in coordinates $(X,r_{\bot})$ under the bounds $x_1=x_2=X$, ${\bf r}_{\bot}\cdot{\bf E}=0$. In the two-electron case the e-e repulsion potential is smoothed at short distances for better visibility of the plot.
}
\label{fig:single_and_collective_potentials}
\end{figure}

Comparing conditions~(\ref{collective_condition}) and~(\ref{sequential_condition}) and analyzing the plots of Fig.1, one can conclude that collective tunneling may dominate if 
\begin{equation}
    \label{CT-ineq}
I_p + V_{\rm ee} < 2I_{p2}~.
\end{equation} 
On one hand, usually $I_p/2<I_{p2}$, on the other, $V_{\rm ee}>0$ and is not negligible in absolute value. 
Inequality (\ref{CT-ineq}), which favors CT, is easier to fulfill in systems with essentially different ionization potentials, e.g. in negative ions or in Li-like positive ions \cite{tyurin-leb23}.  
These qualitative arguments show that CT might potentially be a dominant or at least a non-negligible mechanism of double ionization, but definite conclusions can only be derived on the basis of quantitative analysis.
In the following sections, we present numerical simulations and an analytic theory, which prove the significance of the collective channel.

\section{Statement of the problem and numerical method}

We consider double ionization of an atom in the field of a short intense electromagnetic pulse under the following conditions and constraints:
\begin{enumerate}
    \item Recollision is eliminated. This is achieved by applying an unipolar pulse, which does not allow the outer electron returning back to its parent ion.
    \item In long pulses, the outer electron will be removed on the front edge, so that by the time the field becomes strong enough to disturb the inner electron, the outer orbital can appear depleted.
    Collective ionization requires both electrons to be present in the bound state. Therefore, to reveal the collective channel, short pulses with a steep time envelope, which allow both electrons to survive in the bound states up to the field maximum, look preferable. A qualitative definition of the long- and short-pulse regimes is given in Sec.~\ref{Signatures of collective ionization}.
    \item Numerical TDSE solution for two electrons in 3D requires, for the time intervals necessary to reveal the collective effect, an extremely large computational capacity. As we demonstrate below, in the 1D the collective channel is entirely blocked. Thus, our study is focused mostly on a model 2D two-electron atom, although we also use its 1D analog when appropriate. Because of the restricted dimension, we do not consider numerically effects of angular momentum and focus on a spatially symmetric two-electron state.
    \item Finally, although helium looks the most suitable target for studying double ionization from  a spatially symmetric bound state, its high ionization potentials lead to a necessity of using sub-femtosecond pulses to reach the short-pulse regime (see estimates in Sec.IV). Moreover, the high values of the ionization potentials impose severe requirements on the numerical space and time grids pushing this case to the edge of our computational abilities. For these reasons, we focus our study on a system, which models the ionization potentials of xenon and of the negative bromine ion. The latter is of interest because of a high difference in the two first ionization potentials \cite{becker-prl00,tyurin-leb23}. Analytic results derived in Section V are also applied to helium.
\end{enumerate}

The two-electron TDSE reads
\begin{equation}\label{2eTDSE}
\begin{gathered}
    i\frac{\partial}{\partial t} \Psi(\bm r_1, \bm r_2, t) = \\
    = \big[H_0(\bm r_1, t) + H_0(\bm r_2, t) + V_{\rm ee}(\bm r_1, \bm r_2)\big]\Psi(\bm r_1, \bm r_2, t)~,
\end{gathered}
\end{equation}
where $H_0$ is the single particle Hamiltonian including the interaction with the external electric field
\begin{equation}\label{H0}
    H_{0}(\bm r, t) = -\frac{1}{2} \Delta + V_{\rm ei}(\bm r) - \varphi (\bm r, t)~
\end{equation}
with $\varphi$ being the scalar potential of this interaction. 
The electron-electron interaction $V_{\rm ee}$ and the interaction with the parent ion $V_{\rm ei}$ were approximated by the smoothed Coulomb potentials:
\begin{equation}\label{Vion}
    V_{\rm ei}(\bm r) = \frac{-Z}{\sqrt{\bm r^2 + a^2}}~,
\end{equation}
\begin{equation}\label{Vint}
    V_{\rm ee}(\bm r_1, \bm r_2) = \frac{1}{\sqrt{(\bm r_1 - \bm r_2)^2 + b^2}}~,
\end{equation}
where $Z$ is the residual atomic charge after the removal of two external electrons.
Model potentials of such type are being productively used in numeric studies of atomic systems and their interactions with intense external fields \cite{hall2009energies, fernandez1991analytical,lima1981atoms, silaev2010strong, gebremedhin2014calculations}
The smoothing parameter $a$ is  chosen to reproduce the second ionization potential $I_{p2}$ when the outermost electron is removed.  
After this, the parameter $b$ is adjusted to reproduce the $I_{p1}$ of the outer electron. 
Parameters $a$ and $b$ and the ground-state wavefunction are found by numerically solving the TDSE (the values of $a$ and $b$ are given below). 
The two-electron ground state wave function was obtained by using propagation in imaginary time \cite{bader2013} (not to be confused with the semiclassical imaginary time method in the theory of strong field ionization \cite{popov2005}). 
To suppress recollisions, the time dependence of the electric field ${\bf E}(t)$ linearly polarized along the $x$-axis is set in the form of a soliton-like unipolar pulse with the amplitude value $E_0$ and the full duration $T$ 
\begin{equation}\label{E(t)}
    E(t) = -E_0\sin^2\left(\dfrac{\pi t}{T}\right)~.
\end{equation}
In Eqs. (\ref{2eTDSE}, \ref{H0}), interaction of the electrons with the field of the unipolar pulse is introduced through the scalar potential: $\varphi (\bm r, t) = -\int \tilde E (x, t)dx$, where $\tilde E (x, t)$ is the electric field generally dependent also on coordinate $x$.
In a spatially homogeneous field as (\ref{E(t)}), the electrons are accelerated after ionization, rapidly reaching the boundaries of the computation box.  
This hinders analysis of the ionization dynamics. 
To make such an analysis more affordable, in the numerical calculation, we use the field which is homogeneous in the vicinity of the ion and vanishes far from it:
\begin{equation}\label{E(r,t)}
\tilde E(x, t) = 
   \begin{cases}
       E(t)\cos^2\left(\dfrac{\pi x}{2 x_0}\right),~|x|\leq x_0 \\
       0,~\text{elsewhere}~,
   \end{cases}
\end{equation}
where $x_0=30$ a.u.
With this choice, the field remains almost spatially homogeneous on the scale of the tunnel barrier $I_p/E_0\simeq 10$ a.u. We have verified that increasing $x_0$ up to $300$~a.u., which practically corresponds to a homogeneous field, only affects the post-tunneling acceleration of the electrons, while the ionization dynamics of interest remains weakly sensitive to $x_0$.

The TDSE (\ref{2eTDSE}) is solved for the 1D and 2D systems modeling the neutral xenon atom and negative bromine ion ${\rm Br}^-$, using the split-step Fourier method \cite{weideman1986split}.
For the 1D atom coordinates $x_1$ and $x_2$ of the electrons were discretized on a spatial grid of $2048\times 2048$ nodes with a step $\Delta x = 0.5$ a.u. for ${\rm Br}^-$ and $\Delta x = 0.3$ a.u. for Xe. The time steps are $\Delta t = 0.2$ a.u. for ${\rm Br}^-$ and $\Delta t = 0.1$ a.u. for Xe.
Convergence tests with respect to the size of the spatial and time steps show that the chosen values are sufficiently small to avoid numerical noise, and a further reduction does not affect the results.
The absorbing layer \cite{neuhasuer1989} occupies 15 nodes at the boundary of the computational box. 
The computational box is chosen such that the wave function never reaches the absorbing layer during the calculation. We monitor the norm of the wave function at every time step; as long as it remains equal to unity, no information about double ionization is lost and the results are reliable. Once the norm starts to decay, this signals that the wave packet has reached the absorber, and the corresponding data are discarded. This approach guarantees our results being free from boundary effects.
For the 2D problem, the coordinates were discretized with a step $\Delta x = \Delta y = 0.5$ a.u. and $N_x = 512$, $N_y = 256$ nodes for each electron. 
The total size of the computational box is $512\times256\times512\times256$ nodes. 

To represent the space probability distribution for the two 2D electrons on the plane $(x_1, x_2)$ of the coordinates along the field direction, we integrate $|\Psi({\bf r}_1,{\bf r}_2,t)|^2$ over the transverse coordinates $y_1$ and $y_2$, and anlyze the function 
\begin{equation}\label{F}
    F(x_1, x_2, t) = \int \limits_{
    } |\Psi(x_1, y_1, x_2, y_2, t)|^2 dy_1 dy_2~.
\end{equation}
In (\ref{F}), integration can be performed either over all coordinates, $-\infty<y_{1,2}+\infty$, or with exclusion of an area near the atom.
The latter helps to separate the contributions of single and double ionization.
Indeed, cutting out small values of $y_1$ and $y_2$ eliminates most of the contribution from single ionization (when one of the electrons is located near the origin, $y=0$) and makes the double ionization flux better visible on the background generated by the outer electron.

\section{Signatures of collective ionization}\label{Signatures of collective ionization}

In this Section, we present results of numerical TDSE solutions for neutral xenon and the negative bromine ion and demonstrate signatures of CT in the two-electron ionization flux.

\begin{figure*}[!htb]
\centering
\includegraphics[width=0.8\linewidth]{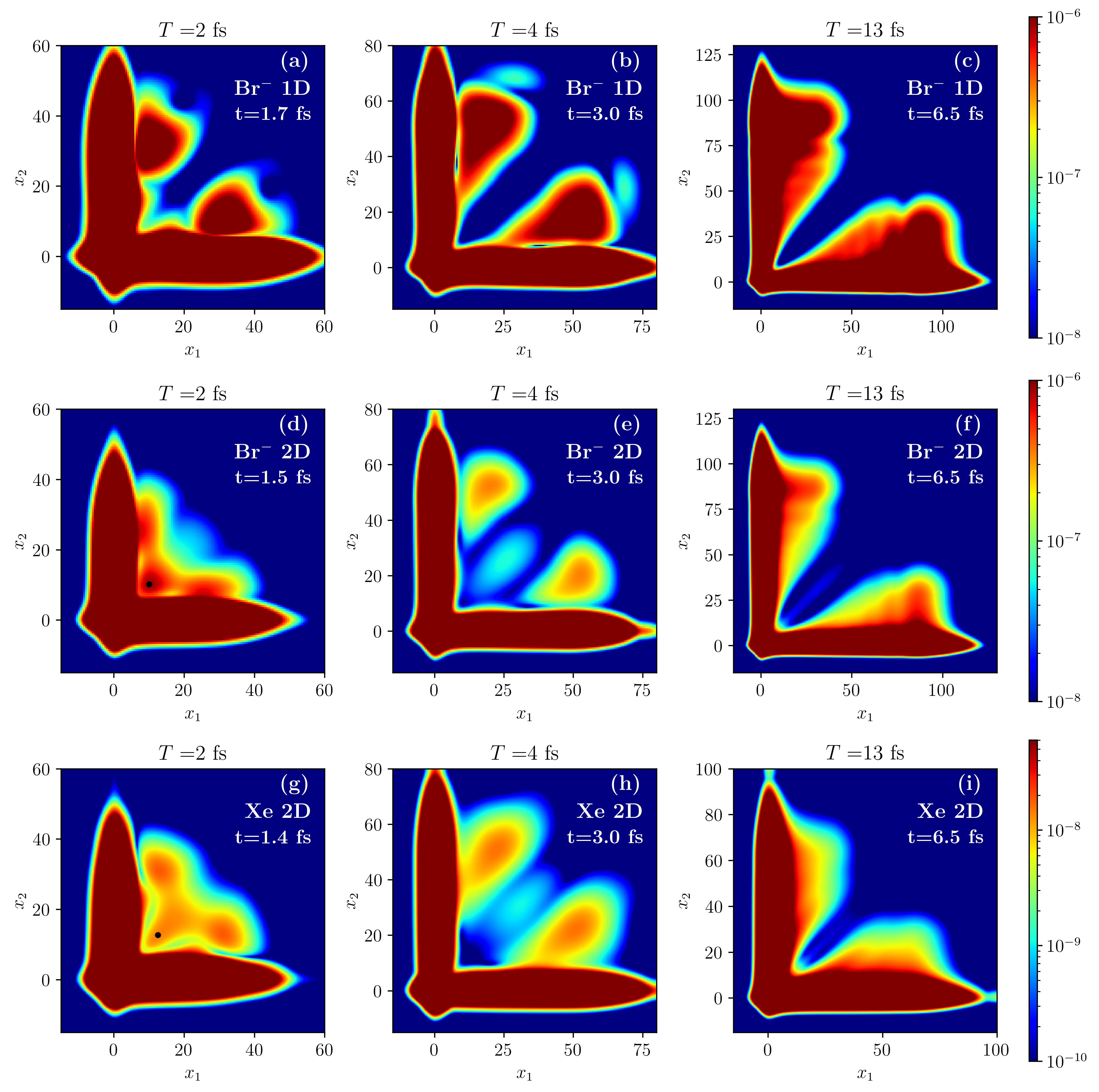}
\caption{Probability density $|\Psi(x_1, x_2, t)|^2$ for the 1D ${\rm Br}^-$ (a--c) and distributions $F(x_1, x_2, t)$ calculated along (\ref{F}) for the 2D ${\rm Br}^-$ (d--f) and 2D Xe (g--i). The field amplitude is $E_0=0.035$ for $\rm{Br}^-$ ($4.3\cdot 10^{13}{\rm W/cm}^2$) and $E_0=0.06$ for Xe ($1.3\cdot 10^{14}{\rm W/cm}^2$). The columns correspond to different pulse durations $T=2$ fs, $T=4$ fs and $T=13$ fs, respectively. 
Black dots on the panels (d) and (g) mark the peaks
of the two-electron ionized wave packet on the diagonal $x_1=x_2$ at $x_{\rm peak}=10$ and $x_{\rm peak}=12$ correspondingly.
}
\label{fig_br:prob_dence_LPR_and_SPR}
\end{figure*}

A necessary condition for collective ionization is that both electrons remain bound until the field reaches the values sufficient to disturb both orbitals. 
Due to a considerable difference in the  ionization potentials, in sufficiently long pulses with a slowly growing field, the outer electron will be removed well before the inner one could be considerably affected by the laser. 
In shorter pulses, the inner electron may begin to feel the field before the outer one is removed.
The border between these two regimes of ionization is determined by the saturation time $t_s$ for the outer electron.
To estimate $t_s$, we determine this time as the pulse duration for which the population $n_1$ of atoms, appearing due to detachment of the outer electron, reaches 0.8 at the field maximum. 
Assuming that the yield of doubly ionized species $n_2$ is much lower than $n_1$, we estimate the population $n_1$ as
\begin{equation}\label{n1}
    n_1(t)=1-\exp\left( -\int_0^t w_1(t')dt' \right)~,
\end{equation}
where $w_1$ is the quasistatic ionization rate for the outer electron.
For actual calculations of the quasistatic single-electron rates, we use the seminal formulas by Smirnov and Chibisov (SC rate) \cite{smirnov_jetp66} and Perelomov, Popov and Terent'yev (PPT rate) \cite{popov_jetp66}.
For the negative bromine ion and for xenon, the estimation gives $t_s \approx 12$ fs for $E_0 = 0.035$; $E_0 = 0.06$, correspondingly.
We use a duration $T = 2$ fs for the shortest pulse, which amply guarantees the short pulse regime for both of these species.
In contrast, the same estimate for helium gives $t_s\approx 3$fs showing that in this case the short pulse regime can only be achieved in the sub-femtosecond domain. 
To illustrate the transition in the ionization dynamics with increasing pulse duration, we also consider $T=4$~fs and $T=13$~fs.
The field amplitudes $E_0=0.035$ and $E_0=0.06$ correspond to the peak intensity $4.3\cdot 10^{13}{\rm W/cm}^2$ and $1.3\cdot 10^{14}{\rm W/cm}^2$.  

Note that the quasistatic rates \cite{smirnov_jetp66,popov_jetp66} used in (\ref{n1}) assume the tunneling regime of ionization specified by the smallness of the Keldysh parameter \cite{keldysh-jetp65}:
\begin{equation}
    \gamma=\frac{\sqrt{2I_p}\omega}{E_0}\ll 1~.
    \label{gamma}
\end{equation}
Here $\omega$ is the effective frequency defined for unipolar pulses as $\omega=\pi/T$ \cite{popov-usp04}.
For all pulse durations used ($T=2,~4\text{ and }13$ fs), the Keldysh parameter falls into the tunnel regimes for the first electron ($\gamma_1 = 0.54,~0.27\text{ and }0.08$ for $T=2,~4\text{ and }13$ fs respectively), which justifies applicability of the PPT rate in the Eq.(\ref{n1}).

Fig.\ref{fig_br:prob_dence_LPR_and_SPR} shows two‑electron probability distributions for $\mathrm{Br}^-$ in 1D (a--c) and 2D (d--f) for three pulse durations: $T=2$~fs, $4$~fs, and $13$~fs. 
To reproduce the first $I_{p1}=3.37$ eV and second $I_{p2}=11.81$ eV ionization potentials of $\rm{Br}^-$ we used $a=1.66$, $b=2.6$ for 1D electrons and $a=1$, $b=2.2$ for 2D electrons in (\ref{Vion}) and (\ref{Vint}). The residual charge in (\ref{Vion}) is $Z=1$.
All the distributions are taken at time instants well after the field maximum, when the ionized flux has already formed.
In the 1D case, the distributions show a clear signature of SI for all pulse durations: the flux, concentrated along the $x_1$ and $x_2$ axes, signals the ionization of the first electron, while the flux directed away from the axes indicates the sequential emission of the second one.
In contrast, the 2D distributions reveal a pronounced flux along the diagonal $x_1 = x_2$, which indicates simultaneous motion of the two electrons along the field direction.
For the shortest pulse $T=2$ fs (panel (d)), this diagonal flux is best visible.
A similar diagonal structure, though less intense, persists at $T=4$ fs (panel (e)).
For the longest pulse $T=13$ fs (panel (f)), the diagonal flux almost entirely vanishes, and the distribution becomes similar to the 1D case (panel (c)), where the sequential channel dominates.
Remarkably, for the 1D electrons (panels (a--c)), the diagonal flux is absent, regardless the pulse duration.

To demonstrate that the diagonal flux is not specific for negative ions, we present our 2D calculations for neutral xenon with $I_{p1}=12.13$~eV and $I_{p2}=21.21$~eV.
The smoothing parameters in Eqs.~(\ref{Vion}) and (\ref{Vint}) are $a=1.546$, $b=2.1$, and the residual charge is $Z=2$.
Fig.\ref{fig_br:prob_dence_LPR_and_SPR}(g--i) shows the corresponding two‑electron probability distributions for the 2D xenon, obtained with the same pulse durations ($T=2,~4\text{ and } 13$) fs for the field amplitude $E_0=0.06$.
The ionization dynamics clearly exhibit a similar diagonal flux as in $\text{Br}^-$.

\begin{figure}[!h]
\centering
\includegraphics[width=0.95\linewidth]{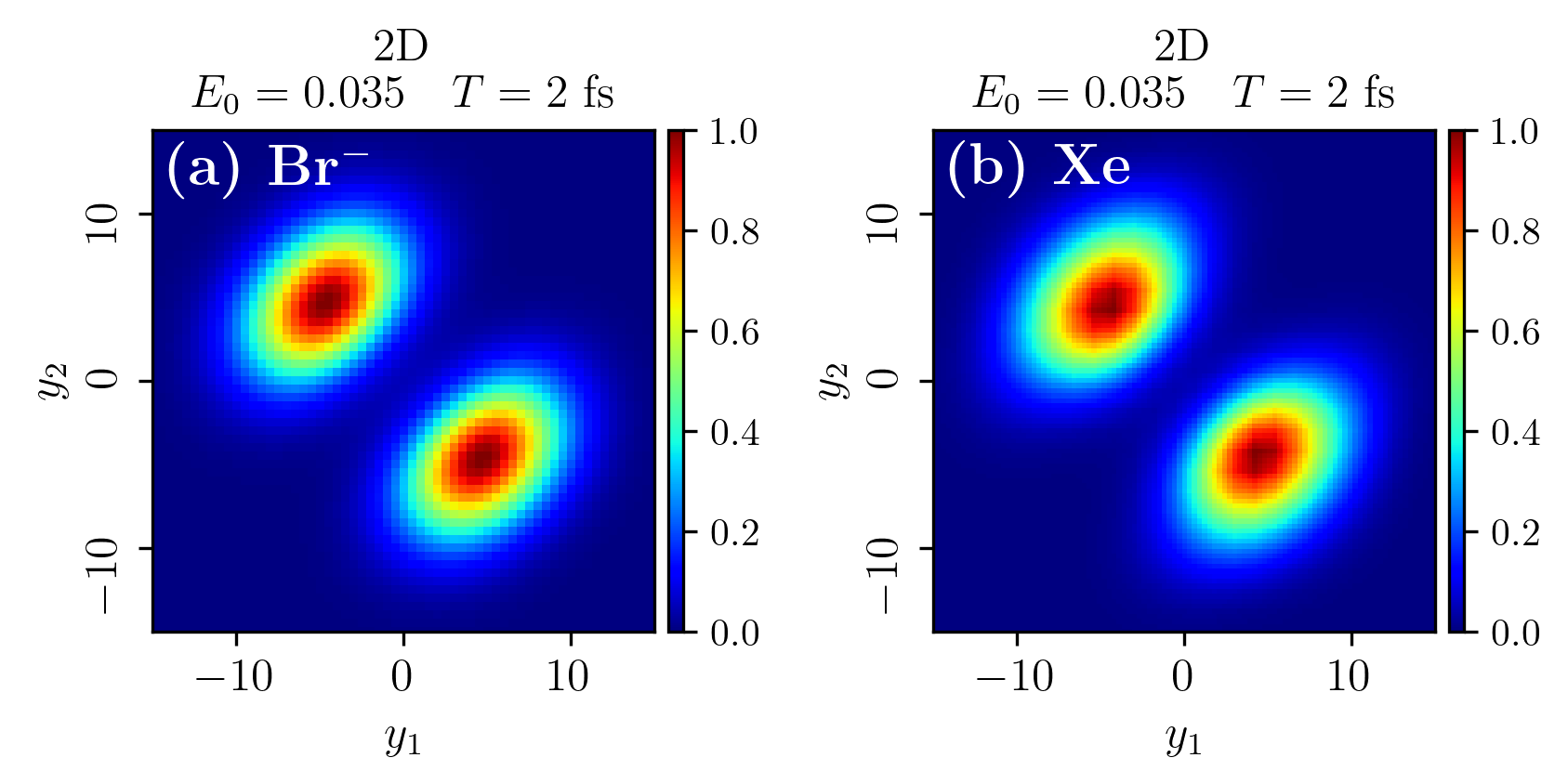}
\caption{
Lateral distributions $|\Psi(x_{\rm peak}, y_1; x_{\rm peak}, y_2)|^2$ normalized to their maximum for $\text{Br}^-$ (a) and Xe (b) at $t=1.5$~fs and $t=1.4$~fs, respectively. 
The values of $x_{\rm peak}$ correspond to the probability maxima on the diagonal $x_1=x_2$ in Fig.\ref{fig_br:prob_dence_LPR_and_SPR}(d, g) marked by the black dots: $x_{\rm peak}=10$ for $\text{Br}^-$ and $x_{\rm peak}=12$ for Xe.
}
\label{fig_br:x1x2y1y2}
\end{figure}

To identify the mechanism responsible for the diagonal flux in the 2D systems, we examine the two-electron motion along the lateral coordinate $y$. 
The distributions of Fig.\ref{fig_br:prob_dence_LPR_and_SPR}(d, g) taken after the field maximum (see the numbers in the figure caption) demonstrate distinct peaks on the diagonal $x_1=x_2$ at $x_{\rm peak}=10$ for $\text{Br}^-$ (d) and $x_{\rm peak}=12$ for Xe (g), i.e., at distances, where the electrons can already be considered as removed from the atom.
The corresponding lateral distribution $|\Psi(x_{\rm peak} , y_1 ; x_{\rm peak} , y_2 )|^2$, exhibits two well separated maxima along the line $y_1 = -y_2$ (Fig.\ref{fig_br:x1x2y1y2}(a, b)). 
This indicates that while the electrons share the same longitudinal coordinate, their most probable transverse coordinates are opposite and essentially nonzero:
\begin{equation}\label{pattern}
    x_1=x_2\equiv x,~y_1=-y_2\equiv y~,
\end{equation}
in agreement with the qualitative description of Section II. 

\begin{figure}[h!]
\centering
\includegraphics[width=1\linewidth]{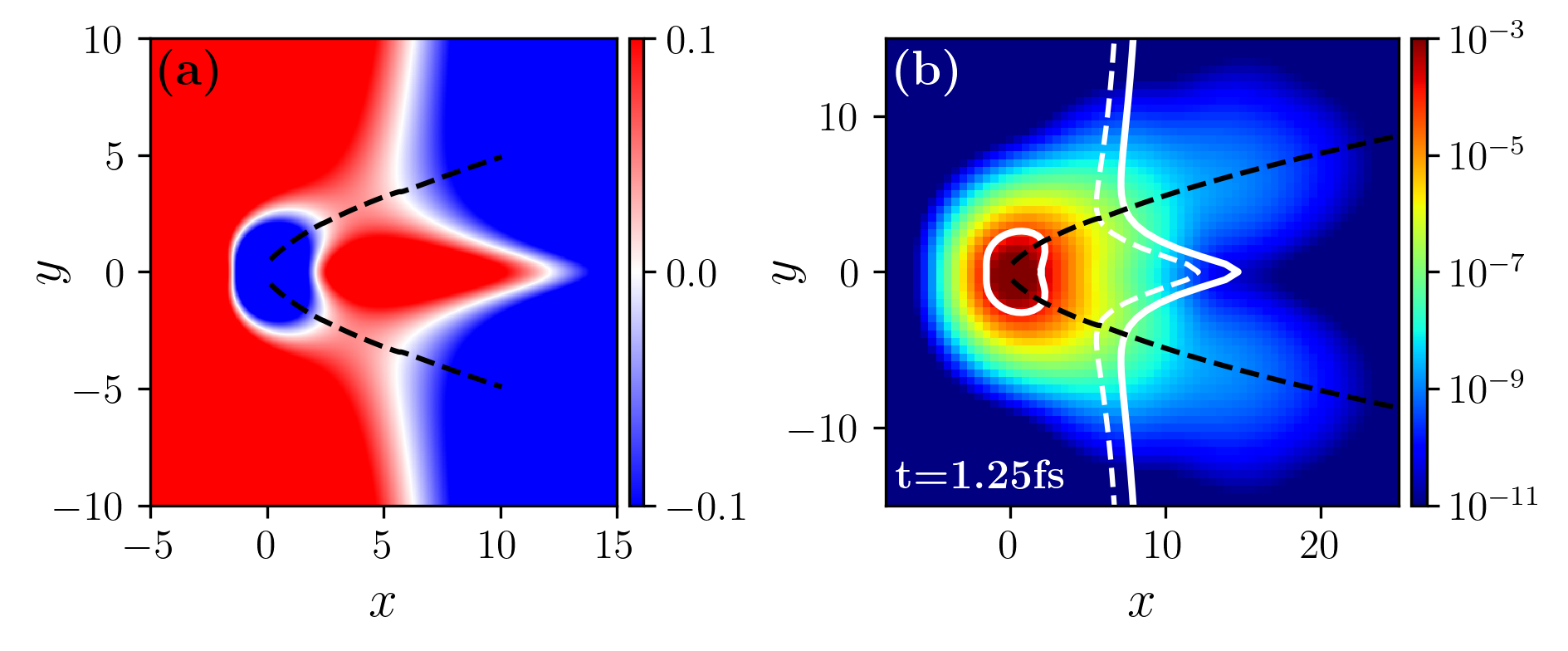}
\caption{Panel (a): the potential surface (\ref{U(x,y)}) for $\text{Br}^-$ at $E=E_0=0.035$, corresponding to the maximum of the electric field at $t=T/2$.
Panel (b): the probability density $|\Psi (x, y;x,-y)|^2$ for $\text{Br}^-$ under condition (\ref{pattern}) at $t=1.25$ fs. 
White solid line shows the potential barrier exit point $x_{\rm exit}(y)$ at $t=1.25$ fs, while the dashed white line gives the same curve at the field maximum $E = E_0$. 
Black dashed curves approximately indicate the trajectories corresponding to the maximum tunneling probability.
}
\label{fig_br:pot_wf}
\end{figure}

This observation, extracted from the \textit{ab initio} calculation, allows for a clear semiclassical interpretation: the field (\ref{E(t)}) drives both electrons jointly along the $x$-axis ($x_1\approx x_2$) while their Coulomb repulsion leads to a lateral separation ($y_1\approx-y_2$).
Correlated motion of this type can potentially result from different physical mechanisms.
If the two-electron potential barrier is suppressed sufficiently to make over-the-barrier escape possible \cite{eckhardt_epj01,eckhardt_pra01,eckhardt_jpb06}, such an ionization channel can be interpreted as direct laser-driven acceleration of both electrons. 
In contrast, if the two electrons moving along symmetrical trajectories can not appear in the continuum without penetrating through the barrier, this channel warrants interpretation in terms of collective tunneling.
To discriminate between the two possibilities, we examine the potential energy (\ref{V_general}) along the trajectories (\ref{pattern}), where it reads:
\begin{equation}\label{V(x,y)}
    V(x, y)=2V_{\rm ei }(x,y)+V_{\rm ee}(x, y; x, -y) - 2E_0 x~.
\end{equation}
This static potential is taken at the maximum of the electric field $E(t=T/2)=E_0$, when tunnel ionization is most probable. 
Then the function
\begin{equation}\label{U(x,y)}
    U(x,y)=V(x,y)+I_p
\end{equation}
is negative in the classically allowed parts of space and positive in the opposite case. 
Fig.\ref{fig_br:pot_wf}(a) shows $U (x, y)$ for the potential of $\text{Br}^-$ and the field amplitude $E_0 = 0.035$. 
The potential barrier separating two classically allowed regions is clearly seen (cf. Fig.\ref{fig:single_and_collective_potentials}(b) above).
The barrier is present even at the field maximum. 
This confirms the tunneling nature of the collective ionization channel associated with the diagonal fluxes shown in Fig.\ref{fig_br:prob_dence_LPR_and_SPR}, and with trajectories (\ref{pattern}). 
 The same behavior was observed for Xe in the pulse with $E_0=0.06$.

The criterion separating collective tunneling and direct laser-driven acceleration can be derived by analyzing the saddles $(x_s,y_s)$ of the function $V(x, y)$ (\ref{V(x,y)}). 
In Fig.\ref{fig_br:pot_wf}(a), their positions are shown by black dots.
Neglecting the smoothing parameters, the solution of $\nabla V(x_s, y_s) = 0$ yields \cite{eckhardt_jpb06}:
\begin{equation}
    x_{\rm s} = \frac{\alpha^3}{2\sqrt{E}}, \qquad
    y_{\rm s} = \pm \frac{\alpha}{2\sqrt{E}},
\end{equation}
where $\alpha = [(4Z)^{2/3}-1]^{1/4}$. 
The barrier-suppression field $E_{\rm BS}$ at which the sub-barrier part of the trajectory vanishes, $V(x_{\rm s}, y_{\rm s}) = -I_p$, is
\begin{equation}\label{eq:E_cr}
    E_{\rm BS} = \frac{I_p^2}{4\alpha^6}.
\end{equation}
For $Z=1$ (negative ions), $\alpha=1.11$ so that $x_s/|y_s|=\alpha^2=1.23$, while for $Z \gg 1$, $\alpha \approx (4Z)^{1/6}$, which makes the most probable trajectory almost 1D.
Consequently, the field (\ref{eq:E_cr}) reduces to the well-known barrier suppression field \cite{popov-usp04,tong_jpb05,kostyukov_pra18} calculated for a quasiparticle with the ionization potential $I_p=I_{p1}+I_{p2}$ and the charge of two elementary charges, $E_{\rm BS}^* = I_p^2/16Z$ \cite{zon-jetp99}.

Thus, for fields $E < E_{\rm BS}$ collective ionization proceeds through tunneling. 
For the $\text{Br}^-$, $E_{\rm BS} \approx 0.041$ exceeds the amplitude value $E_0 = 0.035$ of the pulse used in our calculations. 
For Xe, $E_{\rm BS} \approx 0.072$ is also above the  field amplitude $E_0 = 0.06$, confirming that the diagonal flux observed in both systems indicates CT.

To further substantiate our conclusion, we prove that the ionization flow goes along these optimal paths.
To this end, we examine the wave function's evolution under the condition (\ref{pattern}). 
Fig.\ref{fig_br:pot_wf}(b) shows the distribution $|\Psi(x,y;x,-y)|^2$ at $t=1.25$ fs, just after the field maximum, when the nascent two-electron diagonal flux becomes clearly visible. 
It is seen that the electron density concentrates near the most probable trajectories (black dashed curves in Fig.\ref{fig_br:pot_wf}) passing through the narrowest part of the barrier.
An additional argument in favor of this picture of CT can be made by analyzing the density distribution at earlier time instants, before the electric field is off.
Such distributions, which gain a two-hump shape similar to that shown in Fig.4 for time instants shortly after the interaction has begun, are presented in Appendix \ref{appendix_two_hump}.

Note that the tunneling mechanism revealed above differs considerably from that proposed and described in \cite{zon-jetp99,becker-prl00}, where the electrons were assumed to tunnel as a single quasiparticle, with a full discard of the electron-electron repulsion.
This repulsion non-trivially deforms the potential barrier, whose width appears minimal when the electrons are separated in the lateral direction, $y_1=-y_2$.
Thus, we demonstrated that CT is only possible when the dimension is no less than 2.
Correspondingly, signatures of collective tunneling were absent in the 1D TDSE calculations \cite{lein_prl00}.

\section{Analytic rate of two-electron collective tunneling}

In this Section, we develop an analytic theory that
interprets the {\em ab initio} results and 
allows estimating the rate of two-electron collective tunnel ionization as a function of the parameters of the electromagnetic pulse and the atom.
Our consideration relies on the formalism of the Imaginary Time Method (ITM) \cite{popov_jetp66,popov-usp04,popov2005,poprz_jpb14}.
This method is based on the semi-classical approximation of quantum mechanics and employs classical trajectories solving the Newton equation in complex time and (generally) complex position space.
The ITM has proven itself efficient in the analysis of various phenomena of laser-atom interactions, see e.g. review \cite{poprz_jpb14} and references therein.
Its extensions and generalizations are widely known as the Method of Quantum Orbits \cite{salieres_sci01}, Method of Complex Quantum Trajectories \cite{bauer_book17}, etc.
For the overview of the history and modern developments of this computational technique we send the reader to reviews \cite{poprz_jpb14,bauer_book17,paulus_jpb18,sh-sh_lp25}.
Note that we describe model atomic systems discarding the angular structure of the initial state wave functions. Technically, this allows limiting calculations by the simplest form of ITM, where the ionization probability is found with exponential accuracy.

The starting point of the ITM formalism is a transition amplitude between states $\vert 1\rangle$ and $\vert 2\rangle$
\begin{equation}
    M_{12}\sim\exp(iS_{12})
    \label{M}
\end{equation}
expressed through the classical reduced action 
\begin{equation}
S_{12}=\int\limits_{t_1}^{t_2}(L+\varepsilon_0)dt
\label{S12}
\end{equation}
calculated along a trajectory connecting the initial and final points of system's motion.
Here $L$ is the Lagrangian and $\varepsilon_0$ is the initial energy of the system, $\varepsilon_0=-I_{p1}-I_{p2}\equiv-I_p$ in our case.
The total probability per time unit (rate) reads then
\begin{equation}
    w_{12}={\cal P}\exp(-2{\rm Im}S_{12}).
    \label{w12}
\end{equation}
Eqs.(\ref{M})--(\ref{w12}) hold for the most probable trajectory.
For the extension of these equations to arbitrary trajectories, see \cite{popov-usp04,poprz_jpb14}.
The pre-factor ${\cal P}$ is determined by the structure of the initial and final state wave functions and by the momentum distribution in the final state.
The exponential factor in (\ref{w12}) typically presents the leading ingredient of the tunneling probability, highly sensitive to the atom and field parameters.

For the sake of the following calculations, we notice that:
\begin{itemize}
    \item When ionization below the barrier suppression regime is considered so that 1 in (\ref{M}) means a bound state and 2 corresponds to a free electron with momentum ${\bf p}$, no solutions to the Newton equation exist to connect such initial and final states in real time. Thus, complex-valued (imaginary) time appear in the calculation giving the name to the method.
    \item We develop the theory in the static limit $\gamma\to 0$ (\ref{gamma}), $E=\text{const}$, when all real time instants are equivalent. Therefore, one may set ${\rm Re}(t)=0$ in all calculations and consider time purely imaginary time, $t=i\tau$. This simplifies notations.
    \item The amplitude presented in the form (\ref{M}) can describe also two-particle processes. For this extension, one needs to use the full two-particle Lagrangian with all interactions included.
    \item We calculate the CT  probability with exponential accuracy. Then we only need to consider the most probable trajectory defined by conditions (\ref{pattern}).
\end{itemize}

For the symmetric trajectories  (\ref{pattern}) in imaginary time $t=i\tau$ the action (\ref{S12}) takes the form
\begin{equation}\label{ct:action_ct}
    S = i\int\limits_0^{\tau_0} \left( \dot{x}^2 + \dot{y}^2 + V(x, y) + I_p \right) dt,
\end{equation}
which effectively describes motion of a particle with mass $m_{\rm eff}=2$ in the potential (\ref{V(x,y)}).

We proceed further by applying a set of approximations, which progressively simplify calculations.
Firstly, one can numerically minimize the action (\ref{ct:action_ct}) with the same exact potential (\ref{V(x,y)}) as was used in the numerical TDSE solution.
As a result, we obtain the optimal (most probable) trajectory and the corresponding minimal action $S_{\rm min}^{(1)}$, both depend parametrically on $I_p$, $E$, $a$ and $b$. 
Secondly, the same numerical procedure can be performed with the approximate potential where we neglect the smoothing parameters
\begin{equation}\label{ct:V_eff_total}
    V(x, y) \approx -\dfrac{2Z}{\sqrt{x^2+y^2}} + \dfrac{1}{2|y|} - 2Ex.
\end{equation}
We denote the corresponding minimal action as $S_{\rm min}^{(2)}$.
The applicability conditions of the semi-classical approximation require, in particular, a broad barrier, which means that $\vert S_{\rm min}\vert\gg 1$.
This, in turn, implies that $E<E_{\rm BS}$ (\ref{eq:E_cr}).
When the barrier is sufficiently wide, the dominant contribution to the integral (\ref{ct:action_ct}) comes from a large distance where the ionic potential has its asymptotic Coulomb form, and the same holds for the e-e interaction, because of the significant relative lateral motion of the electrons.
Therefore, we expect that the action is only weakly sensitive to the values of the screening parameters, and $S_{\rm min}^{(1)}\approx S_{\rm min}^{(2)}$.

\begin{figure}[!h]
\centering
\includegraphics[width=1\linewidth]{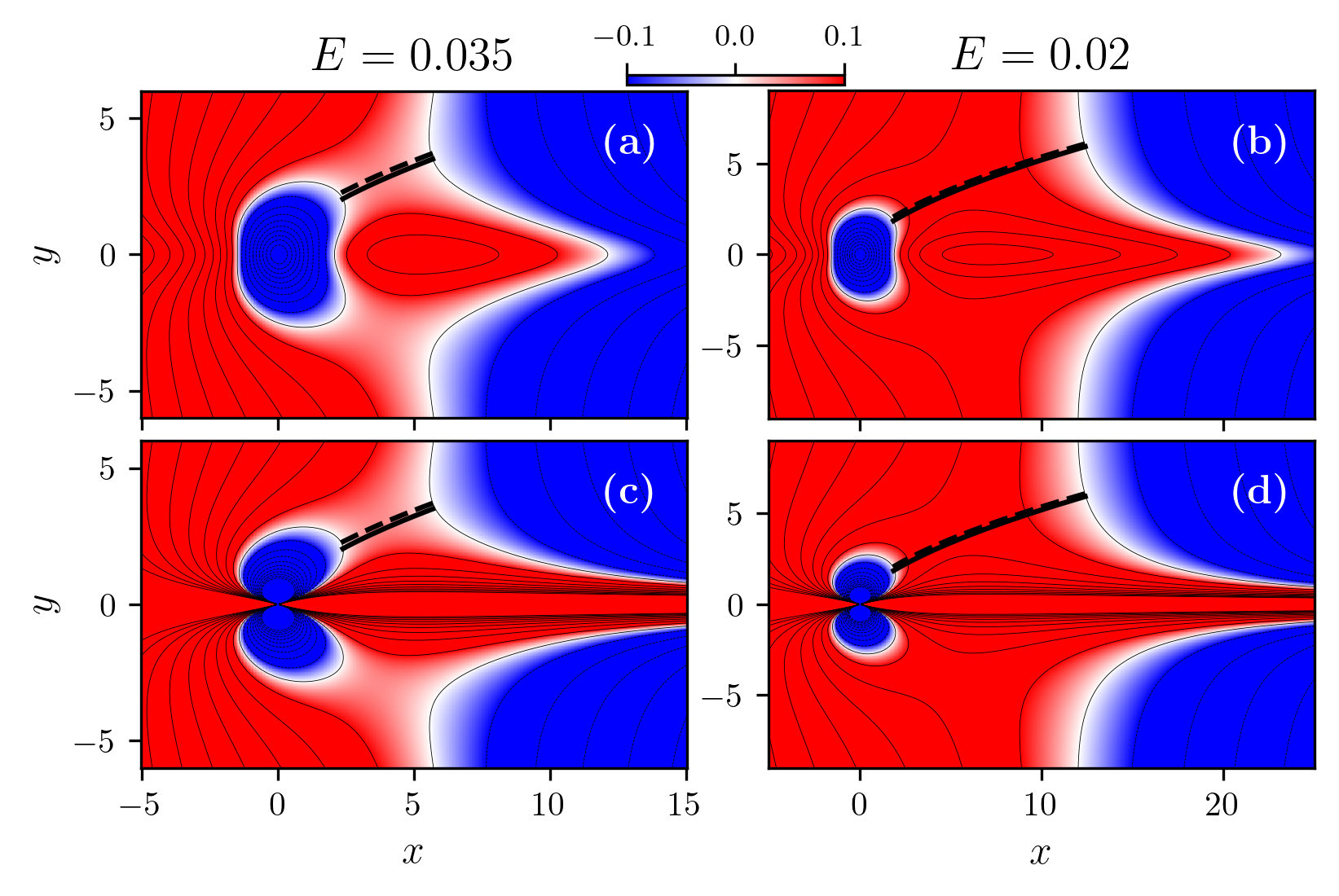}
\caption{
Potential barriers (\ref{U(x,y)}) for the negative bromine ion ${\rm Br}^-$ with $E=0.035$ (a, c) and $E=0.02$ (b, d). Panels (a, b) show the barrier with smoothing parameters $a=1$, $b=2.2$. Panels (c, d) present the same barrier for $a=0$, $b=0$. Black solid and dashed lines show the optimal trajectories in the potential with and without smoothing parameters correspondingly.
The solid and dashed lines appear almost indistinguishable for all presented cases.
}
\label{fig:traj_numeric_1_2}
\end{figure}

In Fig. \ref{fig:traj_numeric_1_2}, the trajectories found from the numeric minimization of (\ref{ct:action_ct}) for ${\rm Br}^-$ in these two cases are shown by black solid and dashed lines correspondingly.
The respective actions are ${\rm Im}S_{\rm min}^{(1)}=1.183$, ${\rm Im}S_{\rm min}^{(2)}=1.228$ for $E=0.035$ and ${\rm Im}S_{\rm min}^{(1)}=7.239$, ${\rm Im}S_{\rm min}^{(2)}=7.246$ for $E=0.02$.
Note that in the first case the inequality $E<E_{\rm BS}$ holds, but the barrier is not actually wide; as a result, the imaginary parts of the actions are of the order of unity.
However, as is typical for the semi-classical regime, the results remain qualitatively correct even near the boundary of the applicability domain, and $\Delta S=\vert {\rm Im}S_{\rm min}^{(1)}-{\rm Im}S_{\rm min}^{(2)}\vert\ll 1$.
In the second case, the barrier width is more than an order of magnitude larger than the size of the initial bound state, leading to a remarkable agreement of the two calculations, $\Delta S=0.017$.

The very fact that the action $S_{\rm min}^{(2)}$ is not only numerically close to $S_{\rm min}^{(1)}$ but finite is not obvious and needs a special comment.
In the ITM applied for description of strong-field ionization, the trajectories start from the origin where the binding state is located \cite{poprz_jpb14}, see Eqs.(\ref{ct:exit_points})-(\ref{ct:trajectory}) in Appendix B.
For the Coulomb potentials $V_{\rm ei}$ and $V_{\rm ee}$ in (\ref{ct:V_eff_total}) free of the smoothing, the integrand in (\ref{ct:action_ct}) diverges at the origin.
For a 1D motion, this leads to a logarithmic divergence known in the theory of the Coulomb correction to the ionization rate \cite{perelomov_jetp67,popov-usp04}.
The regularization of this divergence through matching of the action calculated close to the ion with the asymptotic of the bound-state wave function is one of the key steps in the derivation of single-electron tunnel ionization rates \cite{perelomov_jetp67,popov-usp04,poprz_prl08,poprz_jpb14}.

However, for 2D trajectories, which unavoidably emerge in CT, such divergence does not necessarily occur.
The tiny difference between $S_{\rm min}^{(1)}$ and $S_{\rm min}^{(2)}$ suggests that the new trajectories eliminate the Coulomb divergence and moreover, the contribution of the ion proximity into the sub-barrier imaginary action is small.


As the next step, we derive analytic expressions for the minimal action.
To this end we note that because the optimal trajectories are elongated in the field direction so that $|x|/|y|>1$, interaction of the electrons with the nucleus gives a smaller contribution into the action (\ref{ct:action_ct}) than that between the electrons.
We therefore drop the term $-2Z/\sqrt{x^2+y^2}$ in (\ref{ct:V_eff_total}), so that
\begin{equation}\label{V_short_range}
    V(x,y)\to V_{\rm sr} \approx \dfrac{1}{2|y|} - 2E_0x~.
\end{equation}
This approximation corresponds to ionization of two electrons from a short-range (sr) well and allows to split the action (\ref{ct:action_ct}) in two parts corresponding to motion in the longitudinal ($x$) and lateral ($y$) directions:
\begin{equation}\label{ct:S0=Sx+Sy}
    S^{(3)} = \underbrace{i\int\limits_0^{\tau_0}\bigl(\dot{x}^2 - 2Ex + I_p\bigr)d\tau}_{S_x}~
    \underbrace{\,+~i\int\limits_0^{\tau_0}\left(\dot{y}^2 + \dfrac{1}{2y}\right)d\tau}_{S_y}.
\end{equation}
Details of the calculation of the terms $S_x$ and $S_y$, and the respective trajectories are given in Appendix \ref{appendix_B}. 
In the limit $E\ll E_{\rm BS}$, a set of derivations presented there yields the following compact expressions for the exit point $(x_0, y_0)$, the imaginary time $\tau_0$, and for the action (\ref{ct:S0=Sx+Sy}):
\begin{equation}\label{ct:y_0_approx}
    x_0 \approx \dfrac{I_p}{2E}~,~~y_0 \approx \left( \dfrac{2I_p}{\pi^2 E^2} \right)^{1/3} ~ , ~~\tau_0 \approx \sqrt{\dfrac{I_p}{E^2}}~.
\end{equation}
\begin{equation}\label{ct:S_x}
    S_x\approx i\Bigg\{ \dfrac{2I_p^{3/2}}{3E} 
    \underbrace{- \dfrac{1}{16y_0^2 E \sqrt{I_p}}}_{\Delta S_x} \Bigg\}\equiv S_0+\Delta S_x
\end{equation}

\begin{equation}\label{ct:S_y}
    S_y= i\,\dfrac{3\pi\sqrt{y_0}}{2^{3/2}}.
\end{equation}

In this limit, $x_0$ is just the 1D barrier width for a particle with charge $2e$ and ionization potential $I_p$ in a short-range well.
The ratio $y_0/x_0\simeq E^{1/3}/I_p^{2/3}\ll 1$, which justifies the dominance of the e-e interaction over the attraction to the nucleus as we have assumed above.
The first term in (\ref{ct:S_x}) denoted $S_0$ describes tunneling of the center of mass through the field-created barrier, while the second is a part of the contribution from the electron-electron repulsion.
The second term $\Delta S_x$ vanishes in the limit $y_0 \to \infty$, and $S_x$ reduces to the Zon-Eichmann result \cite{zon-jetp99,becker-prl00}: $S_x \approx i\, 2I_p^{3/2}/(3E)$ obtained under an assumption that the electrons move together, and their mutual repulsion plays no role. 
The ratio of the absolute values of the second term $\Delta S_x=-i/(16y_0^2 E \sqrt{I_p})$  in (\ref{ct:S_x}) and $S_y$ in the limit $E\ll E_{\rm BS}$ can be estimated using (\ref{ct:y_0_approx}) as
\begin{equation}
    \left|\dfrac{\Delta S_x}{S_y}\right|\approx \dfrac{(2\pi)^{3/2}}{48}\left(\dfrac{E}{I_p^2}\right)^{2/3}\ll1~.
\end{equation}
Therefore, $S_y$ provides the dominant contribution to the part of the action generated by the e-e repulsion.

Substituting the approximate expression for $y_0$ from~(\ref{ct:y_0_approx}) and neglecting the term $\Delta S_x$, we obtain the two leading terms of the sub-barrier CT action
\begin{equation}\label{ct:S_short}
    S^{(3)} \approx i\Bigg\{ \dfrac{2I_p^{3/2}}{3E} 
    + \dfrac{3\pi^{2/3}}{2^{4/3}} 
    \left( \dfrac{I_p}{E^2} \right)^{1/6}
   \Bigg\}.
\end{equation}
This formula allows to quantify the role of the e-e repulsion in the collective tunnel effect.
We can estimate the ratio of the second term $S_y$ to the first $S_0$ as
\begin{equation}
    \label{mu}
    \mu=\frac{9\pi^{2/3}}{2^{7/3}}\frac{E^{2/3}}{I_p^{4/3}}.
\end{equation}
With accuracy to a numerical factor $9\pi^{2/3}/2^{7/3}=3.83$ this is the same parameter as (\ref{ct:1/(2y_0 I_p)<<1}) or, equivalently, the ratio $y_0/x_0$. 
In the weak-field limit, $\mu\ll 1$.
For the cases we consider here, $\mu=0.61$ for ${\rm Br}^-$ and $E=0.02$ and $\mu=0.45$ for Xe and $E=0.06$.

In the limit $\mu\ll 1$, which has strictly speaking to be fulfilled for applicability of the semi-classical approximation, the contribution of the electron-electron interaction to the tunneling probability is small compared to that of the electron pair with the external field.
On the other hand, the second term in (\ref{ct:S_short}) is numerically large.
For the same parameters we used above for the estimation of $\mu$, its doubled absolute value is 17.1 and 13.5 correspondingly.
This means that the account of the e-e repulsion is not only crucial for the determination of the most favorable  trajectory for CT, but compulsory for a correct estimation of the tunneling rate.
Without that, this rate appears greatly overestimated, which explains why the results of \cite{zon-jetp99,becker-prl00} were inconsistent with numerical calculations.

As the final step, we account for the attraction of the electron pair to the ion core.
As has been pointed out above, the condition $y_0/x_0\ll 1$ allows to expect that the contribution of this interaction will be smaller in absolute value than that of the e-e repulsion. 
However, it still can appear numerically significant.
The derivation sketched in Appendix 
 \ref{appendix_Coulomb} gives for the Coulomb action
\begin{equation}
\label{SC}
S_C \approx -i\dfrac{2Z}{\sqrt{I_p}}\,
\ln\left( \dfrac{4I_p^{3/2}}{E} \right)~.
\end{equation}
The corresponding Coulomb factor $Q_C=\exp(2iS_C)$ appears in the CT rate.
This leading contribution (see the hierarchy of the applied approximations in Appendix \ref{appendix_Coulomb}) coincides with that found by Zon \cite{zon-jetp99}, while a correction to it due to the lateral electron motion cannot be analytically derived, but is shown to be numerically smaller than (\ref{SC}).

Summarizing, the total CT action and the corresponding rate read
\begin{equation}\label{coulomb:S_CT}
    S_{\rm CT}=S_0+S_y+S_C~,~~~w_{\rm CT}\simeq\exp(-2{\rm Im}S_{\rm CT})
\end{equation}
where $S_0+S_y\equiv S_{\rm sr}$ is defined by Eq.(\ref{ct:S_short}).
The ratio of the Coulomb action (\ref{SC}) to that (\ref{ct:S_y}) induced by the e-e repulsion is also defined by the parameter (\ref{mu}):
\begin{equation}\label{SC/Sy}
    \bigg\vert\frac{S_C}{S_y}\bigg\vert=0.78\frac{E^{1/3}}{I_p^{2/3}}\ln\bigg(\frac{4I_p^{3/2}}{E}\bigg)\sim\sqrt{\mu}\Lambda_{\rm CT}~,
\end{equation}
where we denote $\Lambda_{\rm CT}=\ln(4I_p^{3/2}/E)$ the Coulomb logarithm for collective tunneling, which differs from that for single-electron ionization \cite{popov-usp04} by $\ln(1/\sqrt{2})=-0.35$.

Eqs. (\ref{coulomb:S_CT}) and their ingredients (\ref{ct:S_x}) and (\ref{ct:S_y}) are the central analytic results of this work.
They extend the theory of the atomic tunnel effect \cite{opp_pr28,smirnov_jetp66,perelomov_jetp67} to the case when two electrons penetrate through the barrier simultaneously.
 
Note that the semiclassical theory of tunneling is asymptotically exact for $E\to 0$ \cite{opp_pr28,ankerhold-07,razavy_book13}, and the corresponding formulas for tunneling rates give quantitatively correct results at $E\ll E_{\rm BS}$.
Under these conditions, $\mu\to 0$ (\ref{mu}), and $|S_0|\gg |S_y|\gg |S_C|\gg 1$.
However, for parameters we consider here $\mu\approx 0.5$, and $\Lambda_{\rm CT}=4..5$ so that all three actions appear of comparable value.
This means that the semiclassical approximation in general and the perturbative method used for the calculation of $S_y$ and $S_C$ as corrections to the main action in particular are on the edge of their applicability, but still the results demonstrate a good agreement with those obtained without perturbative expansions as well as with the TDSE solutions.
    
\section{Experimental feasibility: ionization rates and momentum distributions}

In this Section, we analyze the potential experimental feasibility of collective tunneling.
First, let us remind that we study model systems, which reproduce ionization potentials of real atoms, but not the angular structure of the bound state wave functions.
This simplification, customary in the theory of strong field phenomena \cite{keldysh-jetp65,becker-prl00,becker-rmp12,mauger_prl12} including NSDI allows to reproduce such key features as sensitivity of the rate to parameters of the field and atom and the most populated part of the momentum space; the latter determine the overall shape of momentum distributions.
Here we focus on the analysis of these features of CT.
In order to clearly distinguish a contribution of the collective tunnel effect from that of sequential tunneling or some other nonsequential mechanism including recollision, two requirements have to be secured.
\begin{enumerate}
    \item The CT rate has to be at least comparable in absolute value to that of the other competing channels. 
    \item Momentum distribution of the electron pairs or the double charged ions have to bear some unambiguous signatures of the ionization mechanism.
\end{enumerate}
In the following Subsections, we separately address these two points.

\subsection{Comparison of the sequential single-electron and collective ionization rates}

The spatial distribution of the two-electron probability density in 2D systems shown in Fig.1 indicates that, in short pulses the rates of sequential and collective tunneling can be of the same order although the sequential channel remains dominant.
Analytic expressions of the previous section allow to estimate the CT rate with exponential accuracy.
Although single-electron tunneling rates are well known in the literature \cite{smirnov_jetp66,popov_jetp66,perelomov_jetp67,popov-usp04}, comparison of ionization rates is not that straightforward as it could look from the first glance.
Two obstacles for that are: (a) effects of depletion of the outer level and (b) effects of electron-electron repulsion.
As we have shown in a recent paper \cite{tyurin_arxiv25}, the e-e interaction can considerably suppress the sequential ionization rate in short pulses.
From the other side, depletion of the outer electron level on the front edge of a relatively weak laser pulse suppresses CT and favors ST.
As a result, competition of (a) and (b) may considerably influence the CT-ST rivalry depending on the pulse duration and its peak amplitude.
This also prevents, on the present level of the theory, from modeling the competition between CT and ST through classical Monte Carlo trajectory simulations.

\begin{figure}[!h]
    \centering
    \includegraphics[width=0.95\linewidth]{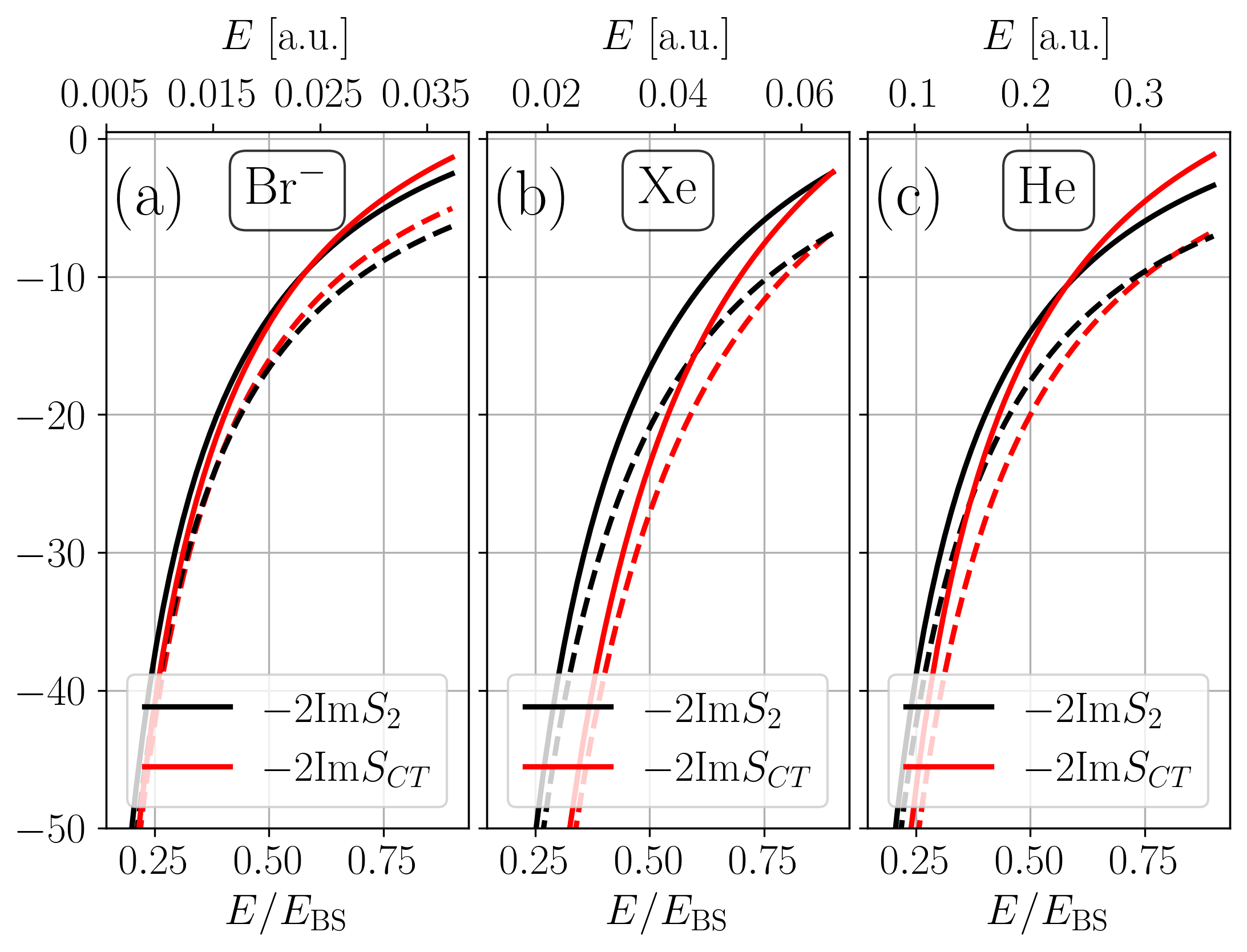}
    \caption{
        Logarithm of the CT ionization rate, $\ln w_{\rm CT}=-2{\rm Im}S_{\rm CT}$ (red lines), and that of the SE rate for the second electron, $\ln w_2=-2{\rm Im}S_2$ (black lines), as functions of the electric field $E$ for ${\rm Br}^-$, Xe, and He. Dashed lines: analytic approximations (\ref{coulomb:S_CT}), (\ref{comparison:S2}); solid lines: numerical results obtained from minimization of the action (\ref{ct:action_ct}) for CT and that for single ionization of the second electron.
    }
    \label{fig:ImS}
\end{figure}

To get a clue to the interplay between collective and sequential tunneling, it is instructive to compare two analytic rates: that of single-electron ionization for the inner electron with ionization potential $I_{p2}$ and the CT rate determined by Eq.(\ref{coulomb:S_CT}).
This comparison helps to estimate the lower bound of the CT contribution in the total DI yield.
We perform this comparison within exponential accuracy with the Coulomb factor accounted for.
Fig.\ref{fig:ImS} shows logarithm of the ionization rate, $\ln w=-2{\rm Im}S$.
For sequential removal of the second electron, 
\begin{equation}\label{comparison:S2}
    {\rm Im} S_2 = 
    \frac{(2I_{p2})^{3/2}}{3E} 
    - \frac{Z}{\sqrt{2I_{p2}}}
    \Lambda_2
\end{equation}
while for CT it is given by (\ref{coulomb:S_CT})
where $\Lambda_2=\ln(2(2I_{p2})^{3/2}/E)$ is the Coulomb logarithm for single-electron tunneling \cite{popov-usp04}.
Dashed lines present the corresponding analytic results (\ref{coulomb:S_CT}) and (\ref{comparison:S2}).
In the weak-field limit, $E\ll E_{\rm BS}$, the analytic and numerical results merge, which is typical for semi-classical asymptotics.
When the field grows, the difference between the analytic approximation and the corresponding numerical results increases gradually.
Note however that for fields $E \simeq E_{\rm BS}$ and above both results become quantitatively incorrect \cite{popov-usp04,tong_jpb05}.

For ${\rm Br}^-$ the CT and single-electron (SE) rates are close in the whole interval of the field strengths.
 This result confirms the suitability of negative ions as systems where the relative contribution of CT can be high in a broad domain of intensities.
For He and particularly Xe, the SE rate dominates that of CT up to $E\approx 0.6E_{\rm BS}$ and $E\approx 0.8E_{\rm BS}$, correspondingly.
Results shown in Fig.1(g,h) for xenon correspond to $E_0/E_{BS}=0.06/0.072=0.83$; the CT and SE rates are also of the same order there.
For He, the CT rate approaches that of ST closer than for Xe. 
However, an estimate of the single-electron tunneling rate for the first electron in He shows that a considerable depletion of the outer orbital can only be avoided for pulses of $\le 1$fs duration, which makes Xe a better experimental candidate for search of CT.
Summarizing, although the results shown in Fig.7 are not unambiguously conclusive, they demonstrate a definite quantitative correlation with the exact TDSE results of Fig.1.
This suggests using our analytic semiclassical results  to estimate parameters optimal for a search of CT in a given atomic species.

\subsection{Signatures of collective tunneling in momentum distributions}

The history of the search for a correct physics interpretation of the NSDI in the 90-es and the followed elaboration of different sub-channels of the recollision mechanism \cite{faria_jmo11,becker-rmp12,bergues_natcom12} proven that momentum distributions of the ions and of the electron pairs carry important and sometimes pivotal information on the correlated ionization dynamics.
In many cases, specific features of such distributions predicted or explained theoretically allowed unambiguously identifying the correlation mechanism.

In this Subsection, we show that for CT there is a certain potentially observable feature of the two-electron momentum distribution, which indicates the presence of this ionization mechanism.
The condition (\ref{pattern}) specifying the most probable trajectory of CT is expected to imprint in the momentum distributions of the electron pairs.
To identify this signature of CT, we study lateral (with respect to the field direction) two-electron distributions in position and momentum space.

\begin{figure}[h!]
\centering
\includegraphics[width=0.98\linewidth]{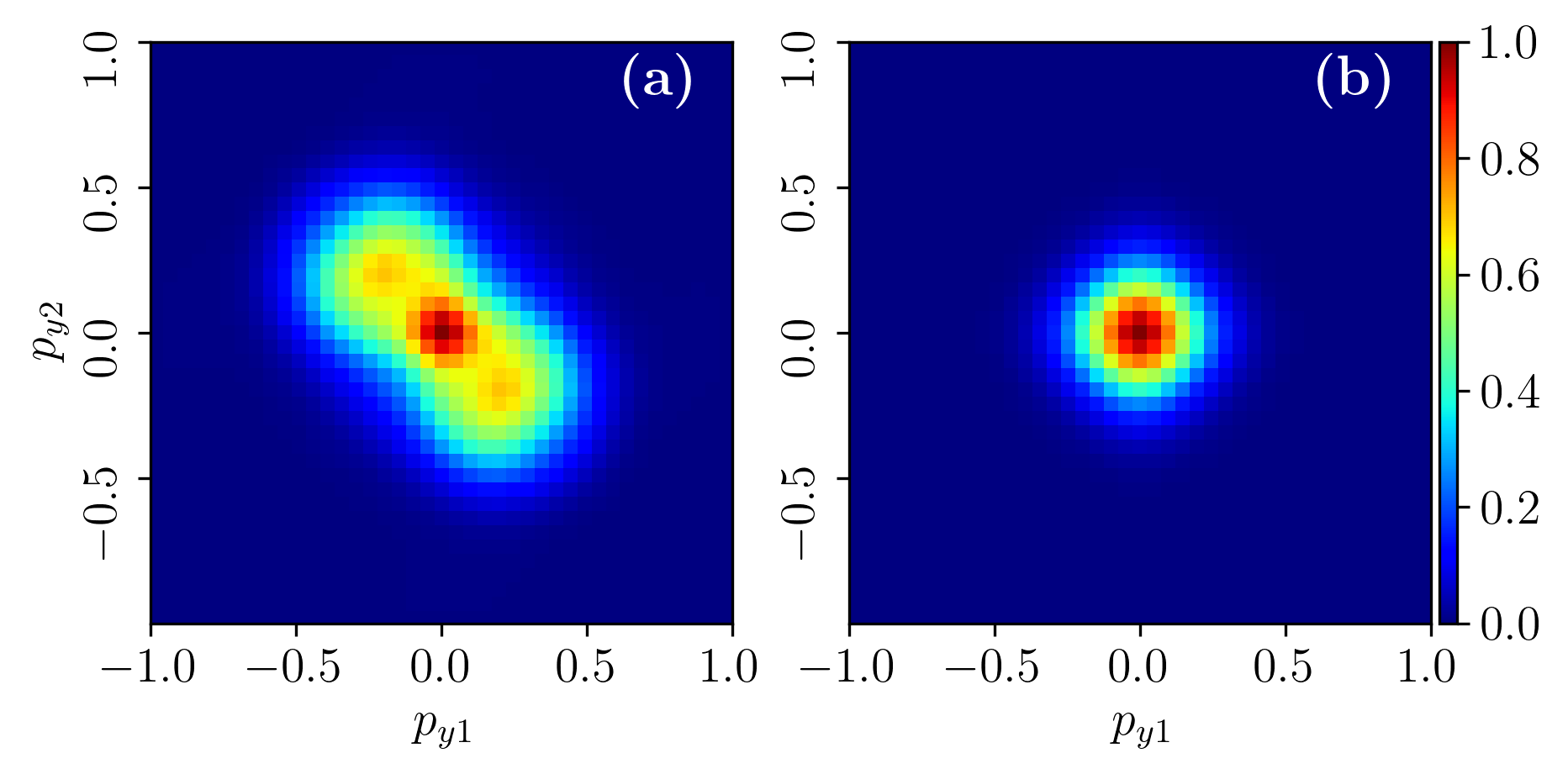}
\caption{Lateral momentum distributions (\ref{md:F_py}) for the electron pair at $t=2.2$ fs, normalized for the maximum values. Panel (a): the distribution for the case with e-e interaction. Panel (b): the distribution of the non-interacting electrons, corresponding to  independent simultaneous double ionization.}
\label{fig:br_y_py}
\end{figure}

To isolate the contribution of double ionization from the total wave function $\Psi(\bm r_1, \bm r_2, t)$, we employ a spatial filter. 
We define $\Psi_{\text{DI}}$ as the part of the wave function, where both electrons are found outside a cutoff distance $r_{\text{cut}} = 15$ a.u. from the ion:
\begin{equation}\label{md:Psi_DI_r1_r2_t}
    \Psi_{\text{DI}}(\mathbf{r}_1, \mathbf{r}_2, t) = \Psi(\mathbf{r}_1, \mathbf{r}_2, t) \eta(|\mathbf{r}_1| - r_{\text{cut}})\eta(|\mathbf{r}_2| - r_{\text{cut}})~.
\end{equation}
Here $\eta(x)$ is the Heaviside function.
This procedure effectively suppresses the contributions from the bound state and from single ionization, where one electron remains close to the nucleus. 
The momentum representation of the wave function is then obtained via the Fourier transform:
\begin{equation}\label{md:Psi_DI_p1_p2_t}
    \Psi_{\text{DI}}(\bm p_1, \bm p_2, t) = \int \Psi_{\text{DI}}(\bm r_1, \bm r_2, t) e^{-i(\bm p_1 \cdot \bm r_1 + \bm p_2 \cdot \bm r_2)}  d\bm r_1 d\bm r_2 .
\end{equation}
To focus on the correlated lateral motion triggered by the electron-electron repulsion, we integrate over the longitudinal momentum components $p_{1x}$ and $p_{2x}$. 
This yields the lateral momentum distribution:
\begin{equation}\label{md:F_py}
    F(p_{1y}, p_{2y}, t) = \int |\Psi_{\text{DI}}(p_{1x}, p_{1y}, p_{2x}, p_{2y}, t)|^2  dp_{1x} dp_{2x}.
\end{equation}
Fig.\ref{fig:br_y_py}(a) presents such a distribution calculated at $t=2.2$ fs for double ionization of ${\rm Br}^-$ at parameters of Fig.1(d) ($T=2$ fs, $E_0=0.035$). 
The distribution exhibits two pronounced side maxima, symmetrically located on the diagonal $p_{1y} = -p_{2y}$. 
These maxima are absent, when double ionization is considered for the same field parameters and in an atom with the same ionization potentials, but without the e-e interaction (Fig.\ref{fig:br_y_py}(b)).
Details of a numerical realization of a two-electron system with non-interacting electrons can be found in Appendix \ref{appendix_noninteract}.

Positions of the side maxima can be quantitatively predicted within the semiclassical model of CT developed above.
The most probable time instant for CT is at the field maximum, $t_0 = T/2 = 1$fs. 
At this time, the two electrons emerge from the tunnel in the correlated configuration defined by (\ref{ct:y_0_approx}) with the initial lateral coordinates $y_1 = -y_2 = y_0(E_0)$ and zero velocity, $\dot{y}_1(t_0) = \dot{y}_2(t_0) = 0$. 
After the exit, the lateral motion in real time is mainly determined by Coulomb repulsion, which dominates, in absolute value, attraction to the nucleus.
In real time, the equation of motion along a trajectory with  $y_1 = -y_2 = y$ reads
\begin{equation}\label{md:real_time_lateral_equation}
    \ddot{y} = \frac{1}{4y^2}, 
\end{equation}
For initial conditions $y(t_0)=y_0(E_0)$, $\dot{y}(t_0)=0$, the solution is (compare with (\ref{ct:trajectory}) in imaginary time)
\begin{equation}\label{md:y_analit}
    t - t_0 = \sqrt{2y_0} \left[ \sqrt{y\,(y - y_0)} + y_0 \, \ln \left( \sqrt{\frac{y}{y_0}} + \sqrt{\frac{y}{y_0} - 1} \right) \right],
\end{equation}
\begin{equation}\label{md:py_analit}
    p_{1y}=-p_{2y}\equiv p_y(t)=\dot{y}(t) = \sqrt{ \frac{1}{2} \left( \frac{1}{y_0} - \frac{1}{y(t)} \right) }~.
\end{equation}

\begin{figure}[h!]
\centering
\includegraphics[width=0.8\linewidth]{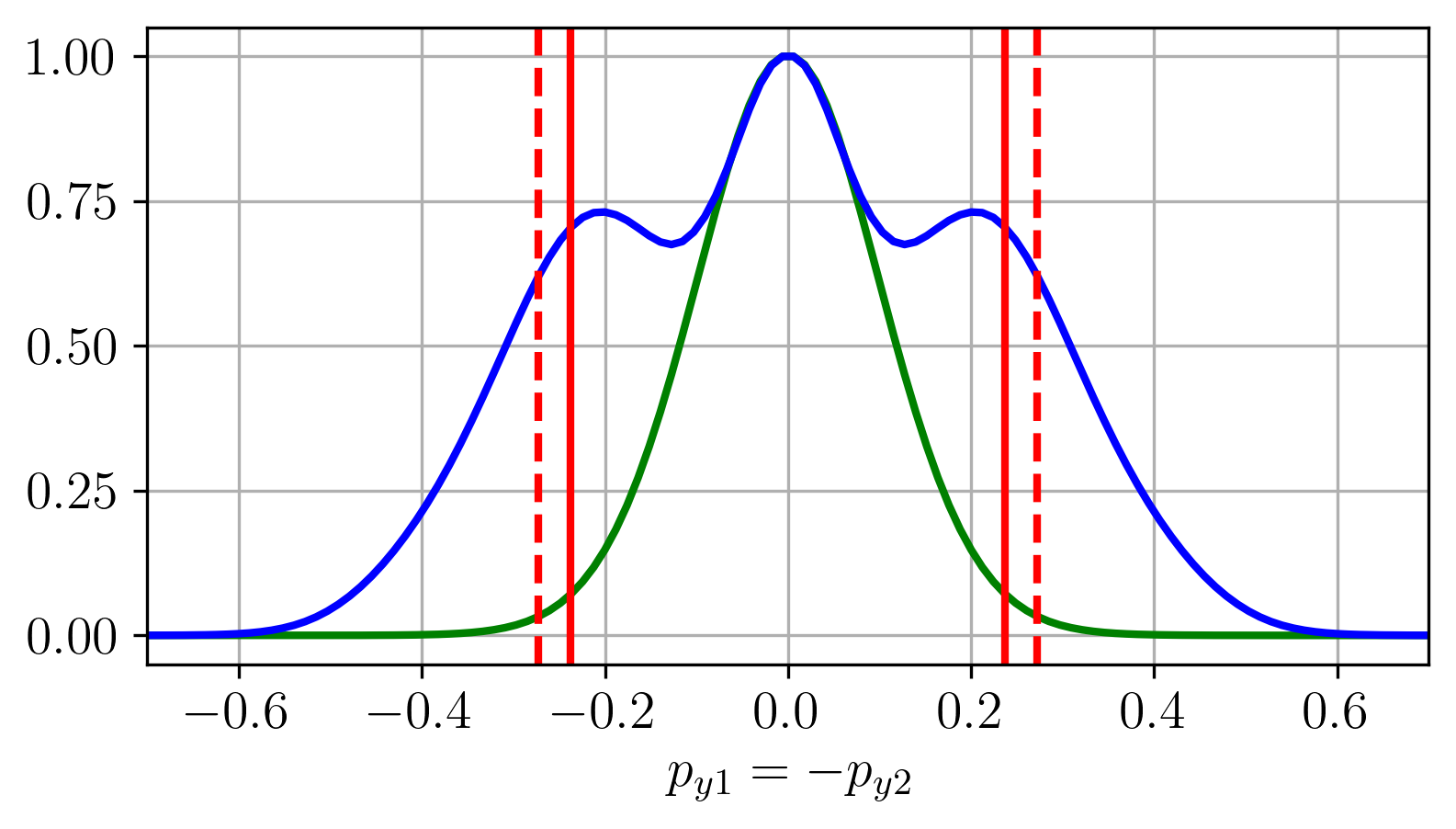}
\caption{Lateral momentum distributions along the diagonal $p_{1y} = -p_{2y}$. 
The blue line corresponds to the interacting electrons (slice of Fig.\ref{fig:br_y_py}(a)); the green line is for the non-interacting case (slice of Fig.\ref{fig:br_y_py}(b)). 
The red dashed line shows the semiclassical prediction for $p_y$ from Eqs.(\ref{md:y_analit}),  (\ref{md:py_analit}). 
The solid red line represents the improved prediction obtained by integrating the equations of motion (\ref{md:full_equation_of_motion}) with an account of the Coulomb focusing effect.}
\label{fig:br_y_py_slice}
\end{figure}

Fig.\ref{fig:br_y_py_slice}(a) shows the diagonal slice $p_{1y}=-p_{2y}$ (blue line) of the lateral momentum distribution of Fig.\ref{fig:br_y_py}(a).
The red dashed line shows the most probable lateral momentum $p_y$, predicted by Eqs. (\ref{md:y_analit}), (\ref{md:py_analit}) for $t = 2.2$ fs. 
Given the simplicity of this analytic model, it looks essentially accurate.
A relatively small, of the order of 20$\%$, discrepancy between the position of the maxima obtained from the TDSE solution and from the analytic estimate, can be attributed to several factors.
\begin{enumerate}
    \item Eq.(\ref{md:real_time_lateral_equation}) neglects the effect of the ion Coulomb potential on the electron dynamics after ionization.
    \item Although the field amplitude $E_0 = 0.035$ satisfies the condition (\ref{eq:E_cr}), it does not fall deeply into the regime of tunneling, $E_0\ll E_{\rm BS}$. 
    As a result, the asymptotic formulas (\ref{ct:y_0_approx}) are not expected to be sufficiently accurate.
    \item
    The analytics above is assumes a constant field, while the slow time dependence of the electric field (\ref{E(t)}) used for the TDSE numerical solution can also affect the momentum distributions.
    \item Finally, the tails of the central peak originating from the sequential ionization channel, overlap with the side structures and effectively pull the lateral maxima toward the origin.
\end{enumerate}

The first factor connected with the  Coulomb focusing effect \cite{brabec_pra96,pepin_jpb05} can be easily accounted for.
To this end, we solve the full equation of motion for the two interacting electrons, which move also under the action of the laser and the Coulomb field of the nucleus. 
For the symmetric trajectory (\ref{pattern}), these equations of motion read
\begin{subnumcases}{\label{md:full_equation_of_motion}} 
\ddot{x} = E-\dfrac{Zx}{(x^2+y^2)^{3/2}}~, \label{md:full_equation_motion_x} \\
\ddot{y} = \dfrac{1}{4y^2}-\dfrac{Zy}{(x^2+y^2)^{3/2}}~. \label{md:full_equation_motion_y}
\end{subnumcases}
By solving them numerically with the initial position (\ref{ct:y_0_approx}) and zero initial velocity we obtain the Coulomb-corrected lateral momentum $p_y =\dot y(t)$. 
The result, shown in Fig.~\ref{fig:br_y_py_slice} by solid red lines clearly demonstrates that the inclusion of the Coulomb focusing considerably improves the quantitative agreement between the TDSE and the semi-classical results.
 
Finally, from Eqs.(\ref{ct:y_0_approx}), (\ref{md:py_analit}) the asymptotic momentum is 
\begin{equation}\label{md:py_limit}
    p_y(t\to\infty)= \left(\dfrac{\pi^2E_0^2}{16I_p}\right)^{1/6}~.    
\end{equation}
This expression estimates the characteristic lateral momentum of the electron pair from the collective channel, predicting the position of the side peaks in the lateral momentum distribution after electron separation.

\begin{figure}[h!]
\centering
\includegraphics[width=0.7\linewidth]{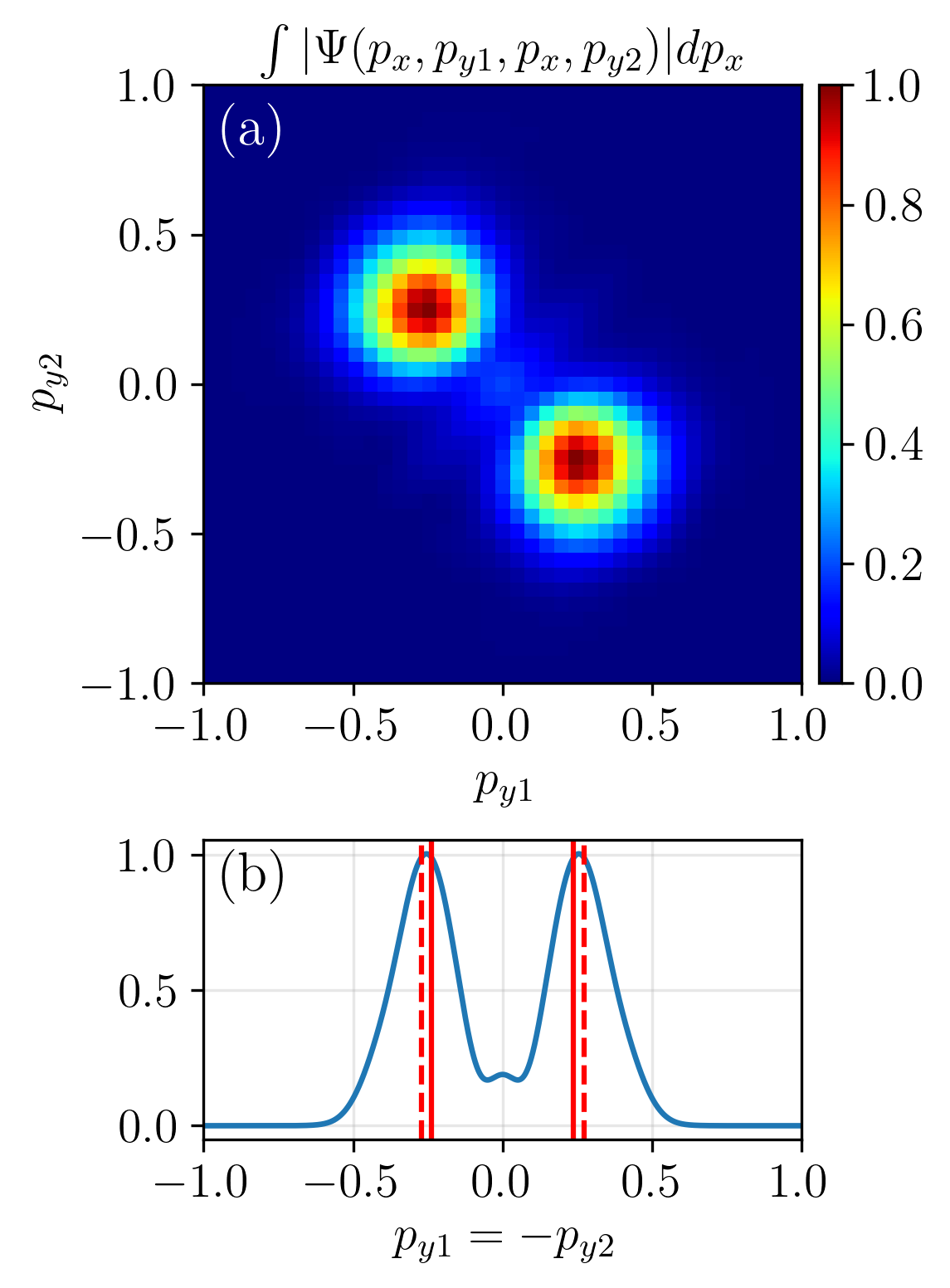}
\caption{Panel (a): normalized lateral momentum distribution integrated over $p_{1x}=p_{2x}$ for ${\rm Br}^-$ at $t=2.2$~fs with interacting electrons. Panel (b): the diagonal slice $p_{1y}=-p_{2y}$ of the panel (a).
The red dashed line shows the semiclassical prediction for $p_y$ from Eqs.(\ref{md:y_analit}),  (\ref{md:py_analit}). 
The solid red line represents the improved prediction obtained by integrating the equations of motion (\ref{md:full_equation_of_motion}) with an account of the Coulomb focusing effect.
}
\label{fig:br_y_py_equal_px}
\end{figure}

To further isolate the contribution of CT, we examine the projection of the two-electron momentum distribution onto the subspace of equal longitudinal momenta, by setting $p_{1x}=p_{2x}$ in (\ref{md:F_py}). 
Such distribution selects the electron pairs that share the same longitudinal momentum, which is precisely the configuration expected for the collective channel, and it therefore provides a better separation of the correlated side peaks from the background produced by sequential ionization. 
The result for ${\rm Br}^-$ at $t=2.2$~fs, with the same field parameters as in Fig.~\ref{fig:br_y_py}, is shown in Fig.~\ref{fig:br_y_py_equal_px} (a). The two side maxima on the diagonal $p_{1y}=-p_{2y}$ appear more pronounced than in the fully integrated distribution of Fig.~\ref{fig:br_y_py}(a).

\begin{figure}[h!]
\centering
\includegraphics[width=1\linewidth]{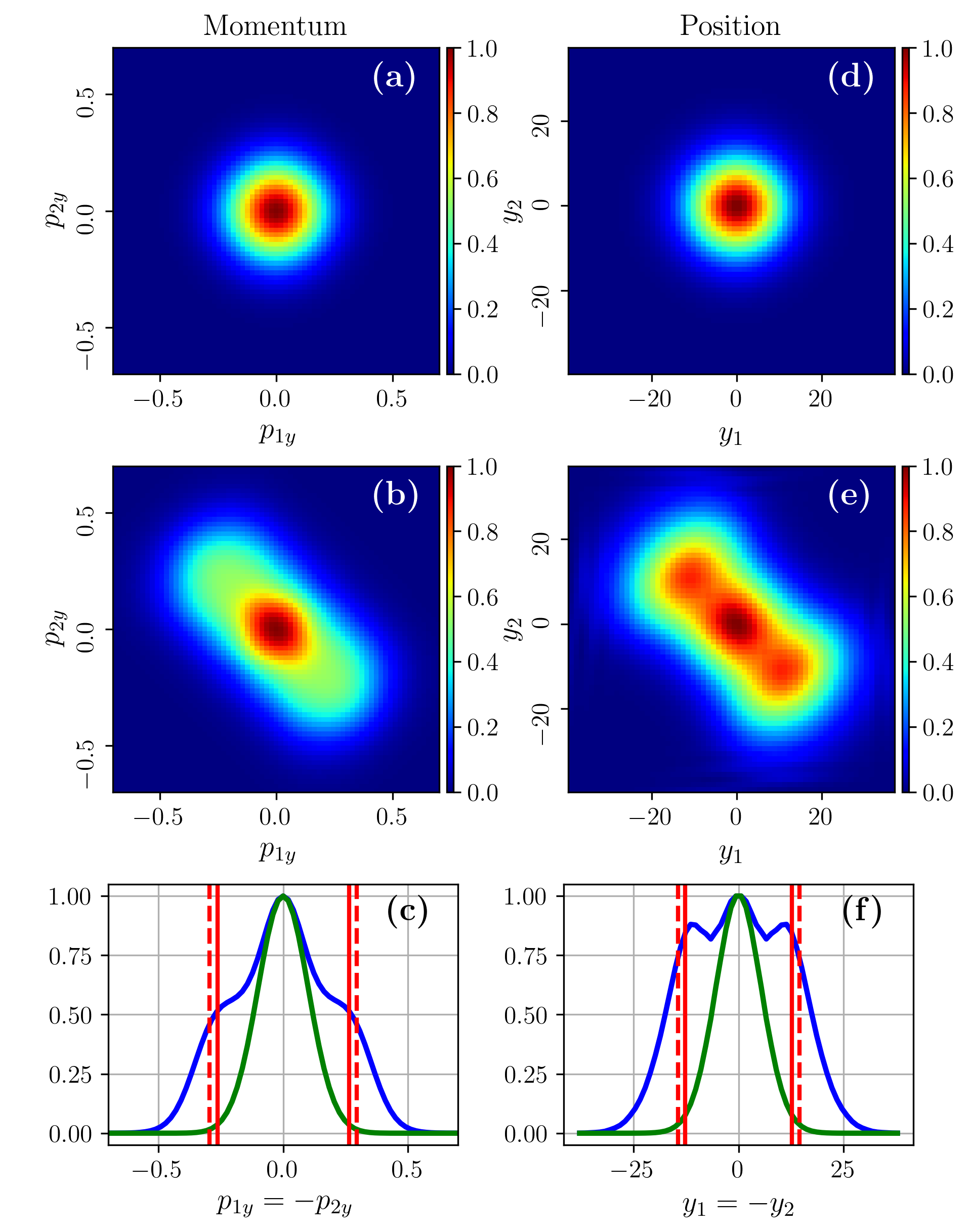}
\caption{Left column: lateral momentum distributions in $(p_{1y},p_{2y})$ (\ref{md:F_py}) for double ionization of Xe ($E_0=0.06,~T=2$ fs). The distributions are recorded at $t=2.2$fs and normalized for the maximum values. 
Panel (a): non-interacting electrons, with the adjusted values of the ionization potentials (see Appendix \ref{appendix_noninteract} for details). Panel (b): interacting electrons. 
Panel (c): the slice along the diagonal $p_{1y} = -p_{2y}$. 
The blue solid line corresponds to the interacting electrons (panel (b)); the green line -- to the non-interacting case (from panel (a)). 
The red dashed line shows the semiclassical prediction for $p_y$ from (Eqs.\ref{md:y_analit}, \ref{md:py_analit}). The red solid  line represents the Coulomb-corrected result (\ref{md:full_equation_of_motion}).
Right column: panels (d,c,f) show the corresponding distributions in position space $(y_1,y_2)$.
}
\label{fig:xe_md}
\end{figure}

Similar lateral electron dynamics takes place in double ionization of neutral atoms if the probability of CT is significant.
Fig.\ref{fig:xe_md} shows an example of Xe, where the relative contribution of CT is lower than in ${\rm Br}^-$, but still not negligible.
In the momentum distributions (b,c), we observe a shoulder-like structure, which considerably broadens the central peak originating due to the sequential process (a).
Remarkably, the three-hump structure survives in the spatial distribution (e,f).
Note that the positions of the side maxima are consistent with the predictions of the semi-classical model (panel (f), red lines).

Summarizing this Subsection, the mutual repulsion of the electrons escaped through CT generates specific double-hump, three-hump, or shoulder-like structures of the lateral momentum distributions.
This structure should become particularly pronounced if the longitudinal momenta are integrated near the line $p_{1x}=p_{2x}$.

\section{Conclusions and outlook}

In conclusion, we presented numeric and analytic results demonstrating the existence and significance of a new and so far unobserved channel of double ionization -- collective tunneling.
We showed that, due to the full discard of the electron-electron interaction, the naive pioneering theory of the collective tunnel effect in atoms \cite{zon-jetp99,becker-prl00} overestimates the rate of this process by a significant factor, which can be as large as several orders of magnitude.
The e-e repulsion accounted for within the semiclassical method, suppresses the CT ionization flow bringing its value to a reasonable agreement with the predictions of exact numerical TDSE solutions.
Thus, the first outcome of our study is that the collective tunnel effect can be present in double ionization of atoms and negative ions at the level comparable to that of the sequential channel.

The significance of the CT rate is necessary but insufficient for experimental verification of this ionization channel.
We showed that the curved 2D sub-barrier motion induced by the e-e repulsion leads to the formation of a three-hump distribution in lateral momenta of the electrons (in case of ${\rm Br}^-$) or at least to its considerable broadening (in case of Xe).
This is an experimentally accessible feature.
{Measurements made under additional bounds imposed on longitudinal momenta can make these signatures of CT much better visible, compared to the integrated 2D distribution (\ref{md:F_py}).
Prediction of these peculiarities in momentum distributions of the electron pairs is the second significant outcome of this work.

However, a considerable gap between the presented theoretical predictions and a realistic experimental scheme targeted at the search for collective tunneling exists. 
To suppress recollisions and make 2D calculations with two interacting electrons attainable, our numerical TDSE solution has been performed for the field of a short unipolar pulse.
Experimentally, there are two short-pulse setups which eliminate recollisions: (a) application of quasi-unipolar extremely short linearly polarized pulses consisting of a single high-field peak and longer low-field shoulders, where the electric field is opposite in sign \cite{arkhipov_jetpl23} and (b) application of short pulses with polarization close to circular (see \cite{litvibyuk_nat19} for an experimental realization and references).
In order to theoretically explore CT in short circularly polarized (CP) pulses and to model momentum distributions, which can be potentially measured in experiments with CP fields, a correct the pre-factor of the momentum-dependent collective tunneling rate has to be found through the quantitative account of (a) the contribution from trajectories other than the most probable and (b) the angular structure of the two-electron bound state wave function.
These tasks are left for the future work.



\section{Acknowledgment}

The authors are thankful to A.V. Flegel, M.V. Frolov, and A.V. Meremyanin for valuable discussions.  
S.V.P. and D.I.T. acknowledge financial support from the Theoretical Physics and Mathematics Advancement Foundation “Basis”.

\appendix
\section{Sub-barrier formation of the two-hump structure}\label{appendix_two_hump}

To confirm that the correlation~(\ref{pattern}) is developing during tunneling rather than after ionization, we examine the wave function in the sub-barrier region. Fig.\ref{fig:y1y2_detailed_evol} shows the transverse distribution for $\text{Br}^-$ at $x_1=x_2=5$, a position under the potential barrier, for the initial bound state (panels (a.1) and (a.2)) and at  $t=1.0$ fs, when the field reaches its maximum (panels (b.1), (b.2)). 
The pronounced maxima along $y_1=-y_2$ (b.1),  (b.2) demonstrate that the correlated configuration is already present in the sub-barrier part of the double ionization dynamics. 
This justifies the relevance of the curved 2D sub-barrier trajectories for the description of the collective ionization pathway.

\begin{figure}[h!]
    \includegraphics[width=1\linewidth]{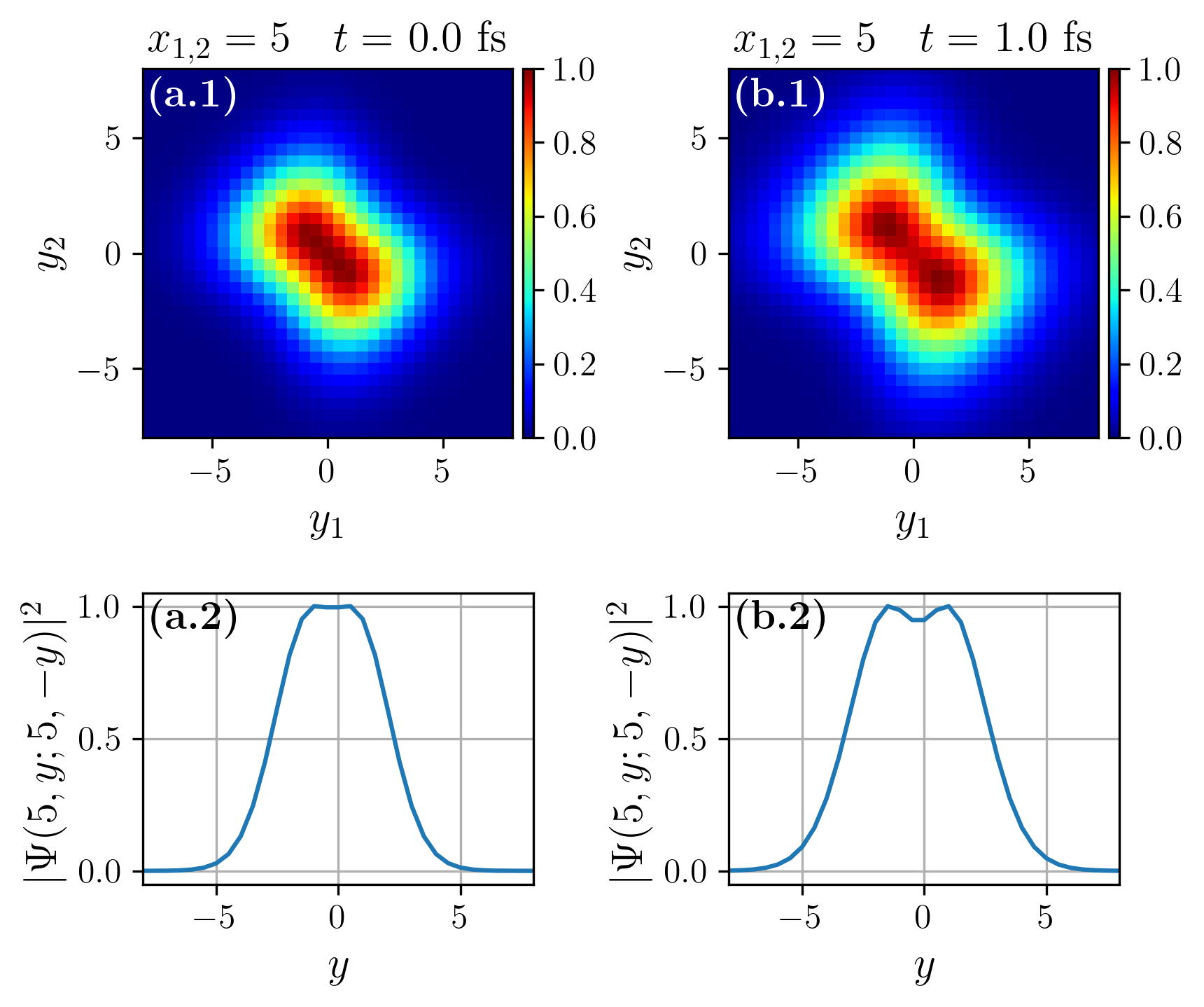}
    \caption{
    (a.1) Transverse distribution $|\Psi(5, y_1; 5, y_2)|^2$ for $\text{Br}^-$ at $t=0$ and $x_1=x_2=5$a.u. (i.e. inside the sub-barrier space domain as it forms after the electric field turns on), normalized to its maximum value. (a.2) Corresponding slice along $y_1=-y_2=y$. Panels (b.1), (b.2) show the same distributions at the field maximum ($t=1$ fs). 
    }
    \label{fig:y1y2_detailed_evol}
\end{figure}

\section{Contribution of the electron-electron interaction to the sub-barrier two-electron action}\label{appendix_B}

In order to find the action (\ref{ct:S0=Sx+Sy}), we need to explicitly calculate the most probable trajectory.
Variations $\delta S_x=0$ and $\delta S_y=0$ lead to the classical equations of motion in imaginary time. 
In the upper half-plane $y>0$, they read
\begin{equation}{\label{ct:equation_of_motion}} 
\ddot{x} = -E,~~~
\ddot{y} = -\dfrac{1}{4y^2}~.
\end{equation}
The exit point $(x_0, y_0)$ is defined by the condition $V_{\rm sr}(x_0, y_0) = -I_p$, which yields the relation:
\begin{equation}\label{ct:exit_points}
    x_0 = \dfrac{I_p}{2E}\left( 1 + \dfrac{1}{2y_0 I_p} \right).
\end{equation}
Taking into account that the kinetic energy vanishes at the exit, we arrive to the boundary conditions for Eqs.(\ref{ct:equation_of_motion}):
\begin{equation}\label{ct:conditions_at_tau=0}
\begin{cases}
    \dot{x}(0) = \dot{y}(0) = 0~,\\
    x(0) = \dfrac{I_p}{2E}\left( 1 + \dfrac{1}{2y_0 I_p} \right)~,\\
    y(0) = y_0~.
\end{cases}
\end{equation}
Here $\tau=0$ is the time instant when the electron pair exits the barrier, while $\tau=\tau_0$ is the initial instant of motion when the electrons are in the atom, ${\bf r}_1={\bf r_2}=0$.
The solution of Eqs.~(\ref{ct:equation_of_motion}) subject to the initial conditions (\ref{ct:conditions_at_tau=0}) reads
\begin{equation}\label{ct:trajectory}
    \begin{cases}
        x(\tau) = -\dfrac{E\tau^2}{2} + \dfrac{I_p}{2E}\left( 1 + \dfrac{1}{2y_0 I_p} \right), \\
        \tau(y) = \sqrt{2}\,y_0^{3/2}\left(
        \arccos\sqrt{\dfrac{y}{y_0}} +
        \sqrt{\dfrac{y}{y_0}}\sqrt{1-\dfrac{y}{y_0}}
        \right).
    \end{cases}
\end{equation}
The parameter $y_0$ is fixed by the requirement that the trajectory reaches the origin simultaneously in $x$ and $y$ directions: $x(\tau_0) = y(\tau_0) = 0$. 
This gives after some algebra

\begin{equation}\label{ct:tau_0}
    \tau_0 = \sqrt{\dfrac{I_p}{E^2}\left( 1 + \dfrac{1}{2y_0 I_p} \right)}.
\end{equation}
\begin{equation}\label{ct:y_0}
    y_0^4 - \dfrac{2y_0I_p}{\pi^2 E^2}\left(1+ \dfrac{1}{2y_0 I_p}\right) = 0,
\end{equation}
Eq.(\ref{ct:y_0}) reveals that $y_0$ scales as $\left(I_p/E^2\right)^{1/3}$, giving
\begin{equation}\label{ct:1/(2y_0 I_p)<<1}
    1/(2y_0 I_p)\sim\left(E/I_p^2\right)^{2/3}\ll1
\end{equation}
in the weak-field limit $E\ll E_{\rm BS}\sim I_p^2$.
Then the leading-order approximation for Eqs.(\ref{ct:exit_points}), (\ref{ct:tau_0}) and (\ref{ct:y_0}) gives Eqs.(\ref{ct:y_0_approx}).

To illustrate the accuracy of our analytics, in Fig. \ref{fig:trajectory} we show the trajectory (\ref{ct:trajectory}) with $y_0$ given by (\ref{ct:y_0_approx}) (green solid line) and that obtained numerically by solving the exact equation (\ref{ct:y_0}) (green dashed line).
Also, the trajectory found by direct numerical minimization of the action (\ref{ct:action_ct}) with the potential energy (\ref{ct:V_eff_total}), which includes the Coulomb interaction with the nucleus is shown by a black line.
These three curves demonstrate close similarity, which confirms that our analytic approximations accurately reproduce the exact tunneling trajectory for weak fields, $E\ll E_{\rm BS}$.

\begin{figure}[h!]
\centering
\includegraphics[width=1\linewidth]{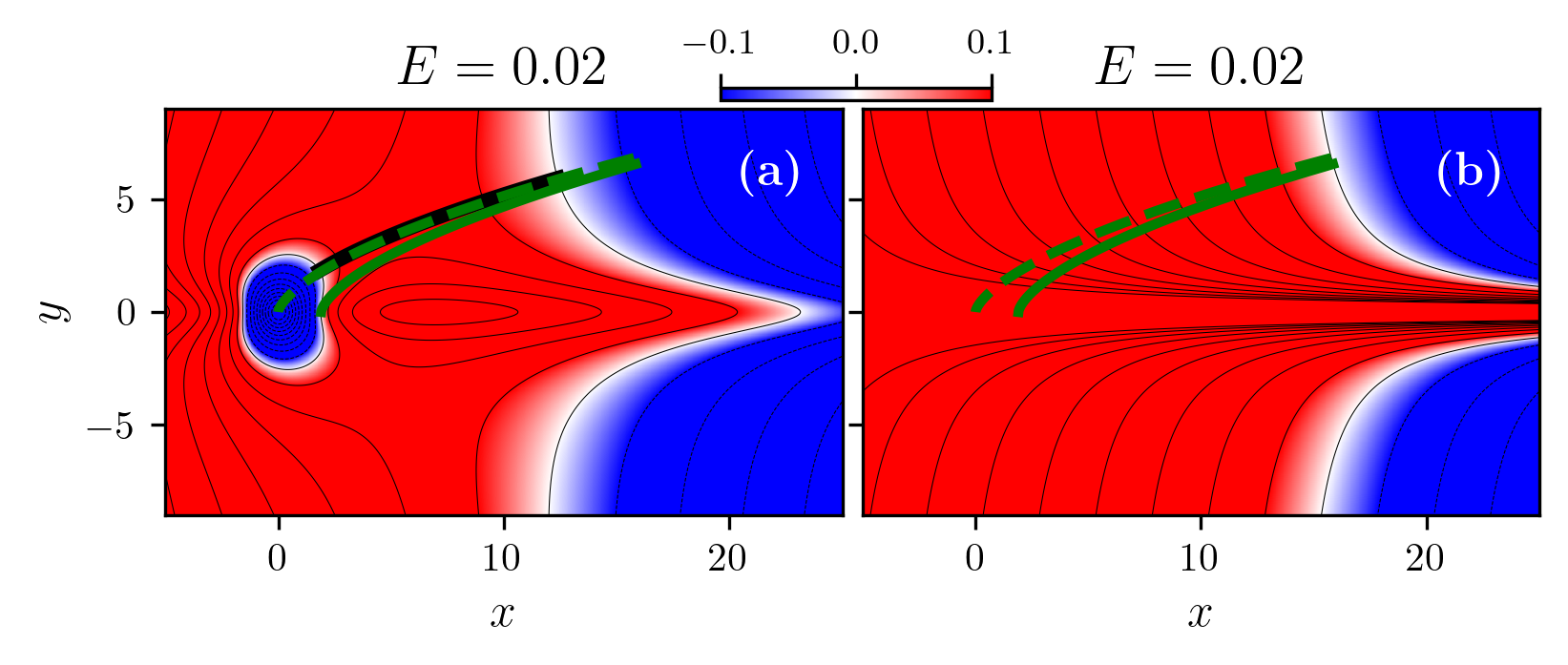}
\caption{Potential barrier $V(x,y)+I_p$ for the negative bromine ion ${\rm Br}^-$ with $E=0.02$ for the full potential (\ref{V(x,y)}) (a) and short-range potential (\ref{V_short_range}) (b). Green lines show the analytic trajectory (\ref{ct:trajectory}) with $y_0$ found numerically from Eq.(\ref{ct:y_0}) (dashed line) and with $y_0$ given by approximation (\ref{ct:y_0_approx}) (solid line). Black line on the panel (a) shows the trajectory found by direct numerical minimization of the action functional (\ref{ct:action_ct}) with the full potential energy (\ref{V(x,y)}).}
\label{fig:trajectory}
\end{figure}

With the trajectory found, we compute the action (\ref{ct:S0=Sx+Sy}). 
The longitudinal contribution $S_x$ can be evaluated using the explicit form of $x(\tau)$ (\ref{ct:trajectory}):
\begin{equation}\label{ct:S_x_accurate}
    S_x = i\left\{ -\dfrac{I_p^{3/2}}{3E}\left( 1+\dfrac{1}{2I_p y_0} \right)^{3/2} + \dfrac{I_p^{3/2}}{E}\sqrt{1+\dfrac{1}{2I_p y_0}}\right\}
\end{equation}
In the weak-field limit, expansion in the small parameter \(1/(2I_p y_0)\ll 1\) (see Eq.~(\ref{ct:1/(2y_0 I_p)<<1})) yields Eq.(\ref{ct:S_x}).

To compute the lateral part $S_y$, we change the integration variable from $\tau$ to $y$ using the first integral of motion  
$\dot{y}^2 = \left( y^{-1} - y_0^{-1} \right)/2$, which follows from (\ref{ct:equation_of_motion}). 
This gives Eq.(\ref{ct:S_y}).

\section{Coulomb factor}\label{appendix_Coulomb}
For single-electron ionization, the contribution of the Coulomb interaction to the total rate and the differential probability has been comprehensively studied both in the tunnel and multiphoton regimes \cite{smirnov_jetp66,perelomov_jetp67,popov-usp04,poprz_prl08}.
The numerically large Coulomb correction $S_C$ to the action generates the corresponding factor $Q_C\gg 1$ in the probability.
Here, we evaluate a similar Coulomb factor for collective tunneling along the most probable 2D trajectory using the semiclassical matching procedure developed for tunneling ionization \cite{poprz_jpb14,popov2005}.

Within the first-order perturbation theory for the action \cite{popov2005}, the Coulomb contribution can be presented as an integral of the electron-ion potential energy along the Coulomb-free sub-barrier trajectory
\begin{equation}\label{coulomb:Sc}
S_C
= -i\int_{0}^{\tau_0}  \frac{2Z\,d\tau}{\sqrt{x(\tau)^2+y(\tau)^2}}~
\end{equation}
with $x(t)$, $y(t)$ and $\tau_0$ defined by Eqs.(\ref{ct:trajectory}) and (\ref{ct:tau_0}), correspondingly.
The key simplification, which allows for an analytic progress employs the different scales of the longitudinal and lateral sub-barrier motions:
\begin{equation}\label{coul:y0x0_ratio}
\frac{y_0}{x_0}\sim\sqrt{\mu}\ll 1
\end{equation}
with $\mu$ given by (\ref{mu}).
Then, far from the ion, $1\ll x<x_0$, the leading term in the integrand is $1/\sqrt{x^2+y^2}\approx1/|x|$, and (\ref{coulomb:Sc}) simplifies to
\begin{equation}\label{Sc_approx}
S_C \approx -2iZ\int_0^{\tau_0}\frac{d\tau}{x(\tau)}.
\end{equation}
Within this approximation, the Coulomb action is just twice that of the single-electron case.
The integral in (\ref{Sc_approx}) diverges logarithmically at $\tau\to\tau_0$ and requires regularization through matching with the asymptotic of the bound state wave function.
As the integral in (\ref{Sc_approx}) is identical to that for single-electron tunneling, the matching procedure can be adopted from the single-electron theory, see \cite{perelomov_jetp67,poprz_jpb14} for details.
This results in (\ref{SC}), see also \cite{zon-jetp99}.

Close to the nucleus, $\sqrt{x^2+y^2}\ll x_0$, the condition $|x|\gg |y|$ inverts and the integrand in (\ref{Sc_approx}) should be replaced by $1/|y(\tau)|$.  
The two asymptotics do not match, but the contribution of that part of space where $|y|\ge |x|$ into the can be estimated by integrating $1/|y(t)|$ from $\tau_0$ to $\tilde{\tau}$ such that $y(\tilde{\tau})=x(\tilde{\tau})$.
This estimation gives for the contribution of this part of the trajectory, $\tilde{S}_C\approx -6iZ/\sqrt{I_p}$, the value small compared to (\ref{SC}) by the factor $1/\Lambda_{\rm CT}$.

\section{Numerical realization for an atom with non-interacting electrons}\label{appendix_noninteract}
In the case of a model atomic system with two non-interacting electrons, the TDSE with the single-particle Hamiltonian (\ref{H0}) was solved for the first and second electrons separately. The two-electron wave function was constructed as the symmetrized product of the single-electron orbitals, corresponding to the full spin $S=0$.

For the negative bromine ion, the potential of the outer electron was approximated by a short-range well of the form
\begin{equation}\label{Vshort}
   V_\text{short}(r) = -c_1\exp\left(-\dfrac{r^2}{c_2^2}\right)
\end{equation}
with two fitting parameters $c_1$ and $c_2$ chosen to reproduce the first ionization potential $I_{p1}=3.37$ eV. 
For the second electron, we used the smoothed Coulomb potential (\ref{Vion}) with a fitting parameter $a$ chosen to reproduce $I_{p2}=11.81$ eV. 
The parameters were taken $c_1=0.48$, $c_2=0.76$ and $a=1.66$ for the 1D problem and $c_1=0.48$, $c_2=2$ and $a=1$ for the 2D problem.

For xenon, the potentials for both electrons were approximated by the smoothed Coulomb potential (\ref{Vion}). 
For the 2D problem, we use $a=1.617$ to reproduce the first ionization potential $I_{p1}=12.13$ eV and $a=2.065$ for the second one $I_{p2}=21.21$ eV.

\bibliography{lit}

\begin{thebibliography}{69}%
\makeatletter
\providecommand \@ifxundefined [1]{%
 \@ifx{#1\undefined}
}%
\providecommand \@ifnum [1]{%
 \ifnum #1\expandafter \@firstoftwo
 \else \expandafter \@secondoftwo
 \fi
}%
\providecommand \@ifx [1]{%
 \ifx #1\expandafter \@firstoftwo
 \else \expandafter \@secondoftwo
 \fi
}%
\providecommand \natexlab [1]{#1}%
\providecommand \enquote  [1]{``#1''}%
\providecommand \bibnamefont  [1]{#1}%
\providecommand \bibfnamefont [1]{#1}%
\providecommand \citenamefont [1]{#1}%
\providecommand \href@noop [0]{\@secondoftwo}%
\providecommand \href [0]{\begingroup \@sanitize@url \@href}%
\providecommand \@href[1]{\@@startlink{#1}\@@href}%
\providecommand \@@href[1]{\endgroup#1\@@endlink}%
\providecommand \@sanitize@url [0]{\catcode `\\12\catcode `\$12\catcode
  `\&12\catcode `\#12\catcode `\^12\catcode `\_12\catcode `\%12\relax}%
\providecommand \@@startlink[1]{}%
\providecommand \@@endlink[0]{}%
\providecommand \url  [0]{\begingroup\@sanitize@url \@url }%
\providecommand \@url [1]{\endgroup\@href {#1}{\urlprefix }}%
\providecommand \urlprefix  [0]{URL }%
\providecommand \Eprint [0]{\href }%
\providecommand \doibase [0]{http://dx.doi.org/}%
\providecommand \selectlanguage [0]{\@gobble}%
\providecommand \bibinfo  [0]{\@secondoftwo}%
\providecommand \bibfield  [0]{\@secondoftwo}%
\providecommand \translation [1]{[#1]}%
\providecommand \BibitemOpen [0]{}%
\providecommand \bibitemStop [0]{}%
\providecommand \bibitemNoStop [0]{.\EOS\space}%
\providecommand \EOS [0]{\spacefactor3000\relax}%
\providecommand \BibitemShut  [1]{\csname bibitem#1\endcsname}%
\let\auto@bib@innerbib\@empty
\bibitem [{\citenamefont {Suran}\ and\ \citenamefont
  {Zapesochny}(1975)}]{suran_jtp75}%
  \BibitemOpen
  \bibfield  {author} {\bibinfo {author} {\bibfnamefont {V.}~\bibnamefont
  {Suran}}\ and\ \bibinfo {author} {\bibfnamefont {I.}~\bibnamefont
  {Zapesochny}},\ }\href@noop {} {\bibfield  {journal} {\bibinfo  {journal}
  {Soviet Technical Physics Letters}\ }\textbf {\bibinfo {volume} {1}},\
  \bibinfo {pages} {2} (\bibinfo {year} {1975})}\BibitemShut {NoStop}%
\bibitem [{\citenamefont {l'Huillier}\ \emph {et~al.}(1983)\citenamefont
  {l'Huillier}, \citenamefont {Lompre}, \citenamefont {Mainfray},\ and\
  \citenamefont {Manus}}]{huillier_pra83}%
  \BibitemOpen
  \bibfield  {author} {\bibinfo {author} {\bibfnamefont {A.}~\bibnamefont
  {l'Huillier}}, \bibinfo {author} {\bibfnamefont {L.}~\bibnamefont {Lompre}},
  \bibinfo {author} {\bibfnamefont {G.}~\bibnamefont {Mainfray}}, \ and\
  \bibinfo {author} {\bibfnamefont {C.}~\bibnamefont {Manus}},\ }\href@noop {}
  {\bibfield  {journal} {\bibinfo  {journal} {Physical Review A}\ }\textbf
  {\bibinfo {volume} {27}},\ \bibinfo {pages} {2503} (\bibinfo {year}
  {1983})}\BibitemShut {NoStop}%
\bibitem [{\citenamefont {Fittinghoff}\ \emph {et~al.}(1992)\citenamefont
  {Fittinghoff}, \citenamefont {Bolton}, \citenamefont {Chang},\ and\
  \citenamefont {Kulander}}]{kulander_prl92}%
  \BibitemOpen
  \bibfield  {author} {\bibinfo {author} {\bibfnamefont {D.~N.}\ \bibnamefont
  {Fittinghoff}}, \bibinfo {author} {\bibfnamefont {P.~R.}\ \bibnamefont
  {Bolton}}, \bibinfo {author} {\bibfnamefont {B.}~\bibnamefont {Chang}}, \
  and\ \bibinfo {author} {\bibfnamefont {K.~C.}\ \bibnamefont {Kulander}},\
  }\href@noop {} {\bibfield  {journal} {\bibinfo  {journal} {Physical review
  letters}\ }\textbf {\bibinfo {volume} {69}},\ \bibinfo {pages} {2642}
  (\bibinfo {year} {1992})}\BibitemShut {NoStop}%
\bibitem [{\citenamefont {Walker}\ \emph {et~al.}(1994)\citenamefont {Walker},
  \citenamefont {Sheehy}, \citenamefont {DiMauro}, \citenamefont {Agostini},
  \citenamefont {Schafer},\ and\ \citenamefont {Kulander}}]{walker_prl94}%
  \BibitemOpen
  \bibfield  {author} {\bibinfo {author} {\bibfnamefont {B.}~\bibnamefont
  {Walker}}, \bibinfo {author} {\bibfnamefont {B.}~\bibnamefont {Sheehy}},
  \bibinfo {author} {\bibfnamefont {L.~F.}\ \bibnamefont {DiMauro}}, \bibinfo
  {author} {\bibfnamefont {P.}~\bibnamefont {Agostini}}, \bibinfo {author}
  {\bibfnamefont {K.~J.}\ \bibnamefont {Schafer}}, \ and\ \bibinfo {author}
  {\bibfnamefont {K.~C.}\ \bibnamefont {Kulander}},\ }\href@noop {} {\bibfield
  {journal} {\bibinfo  {journal} {Physical review letters}\ }\textbf {\bibinfo
  {volume} {73}},\ \bibinfo {pages} {1227} (\bibinfo {year}
  {1994})}\BibitemShut {NoStop}%
\bibitem [{\citenamefont {Kondo}\ \emph {et~al.}(1993)\citenamefont {Kondo},
  \citenamefont {Sagisaka}, \citenamefont {Tamida}, \citenamefont {Nabekawa},\
  and\ \citenamefont {Watanabe}}]{kondo_pra93}%
  \BibitemOpen
  \bibfield  {author} {\bibinfo {author} {\bibfnamefont {K.}~\bibnamefont
  {Kondo}}, \bibinfo {author} {\bibfnamefont {A.}~\bibnamefont {Sagisaka}},
  \bibinfo {author} {\bibfnamefont {T.}~\bibnamefont {Tamida}}, \bibinfo
  {author} {\bibfnamefont {Y.}~\bibnamefont {Nabekawa}}, \ and\ \bibinfo
  {author} {\bibfnamefont {S.}~\bibnamefont {Watanabe}},\ }\href@noop {}
  {\bibfield  {journal} {\bibinfo  {journal} {Physical Review A}\ }\textbf
  {\bibinfo {volume} {48}},\ \bibinfo {pages} {R2531} (\bibinfo {year}
  {1993})}\BibitemShut {NoStop}%
\bibitem [{\citenamefont {Cornaggia}\ and\ \citenamefont
  {Hering}(1998)}]{cornaggia_jpb98}%
  \BibitemOpen
  \bibfield  {author} {\bibinfo {author} {\bibfnamefont {C.}~\bibnamefont
  {Cornaggia}}\ and\ \bibinfo {author} {\bibfnamefont {P.}~\bibnamefont
  {Hering}},\ }\href@noop {} {\bibfield  {journal} {\bibinfo  {journal}
  {Journal of Physics B: Atomic, Molecular and Optical Physics}\ }\textbf
  {\bibinfo {volume} {31}},\ \bibinfo {pages} {L503} (\bibinfo {year}
  {1998})}\BibitemShut {NoStop}%
\bibitem [{\citenamefont {Weber}\ \emph {et~al.}(2000)\citenamefont {Weber},
  \citenamefont {Weckenbrock}, \citenamefont {Staudte}, \citenamefont
  {Spielberger}, \citenamefont {Jagutzki}, \citenamefont {Mergel},
  \citenamefont {Afaneh}, \citenamefont {Urbasch}, \citenamefont {Vollmer},
  \citenamefont {Giessen} \emph {et~al.}}]{weber_prl00}%
  \BibitemOpen
  \bibfield  {author} {\bibinfo {author} {\bibfnamefont {T.}~\bibnamefont
  {Weber}}, \bibinfo {author} {\bibfnamefont {M.}~\bibnamefont {Weckenbrock}},
  \bibinfo {author} {\bibfnamefont {A.}~\bibnamefont {Staudte}}, \bibinfo
  {author} {\bibfnamefont {L.}~\bibnamefont {Spielberger}}, \bibinfo {author}
  {\bibfnamefont {O.}~\bibnamefont {Jagutzki}}, \bibinfo {author}
  {\bibfnamefont {V.}~\bibnamefont {Mergel}}, \bibinfo {author} {\bibfnamefont
  {F.}~\bibnamefont {Afaneh}}, \bibinfo {author} {\bibfnamefont
  {G.}~\bibnamefont {Urbasch}}, \bibinfo {author} {\bibfnamefont
  {M.}~\bibnamefont {Vollmer}}, \bibinfo {author} {\bibfnamefont
  {H.}~\bibnamefont {Giessen}},  \emph {et~al.},\ }\href@noop {} {\bibfield
  {journal} {\bibinfo  {journal} {Physical review letters}\ }\textbf {\bibinfo
  {volume} {84}},\ \bibinfo {pages} {443} (\bibinfo {year} {2000})}\BibitemShut
  {NoStop}%
\bibitem [{\citenamefont {Moshammer}\ \emph {et~al.}(2000)\citenamefont
  {Moshammer}, \citenamefont {Feuerstein}, \citenamefont {Schmitt},
  \citenamefont {Dorn}, \citenamefont {Schr{\"o}ter}, \citenamefont {Ullrich},
  \citenamefont {Rottke}, \citenamefont {Trump}, \citenamefont {Wittmann},
  \citenamefont {Korn} \emph {et~al.}}]{moshammer_prl00}%
  \BibitemOpen
  \bibfield  {author} {\bibinfo {author} {\bibfnamefont {R.}~\bibnamefont
  {Moshammer}}, \bibinfo {author} {\bibfnamefont {B.}~\bibnamefont
  {Feuerstein}}, \bibinfo {author} {\bibfnamefont {W.}~\bibnamefont {Schmitt}},
  \bibinfo {author} {\bibfnamefont {A.}~\bibnamefont {Dorn}}, \bibinfo {author}
  {\bibfnamefont {C.~D.}\ \bibnamefont {Schr{\"o}ter}}, \bibinfo {author}
  {\bibfnamefont {J.}~\bibnamefont {Ullrich}}, \bibinfo {author} {\bibfnamefont
  {H.}~\bibnamefont {Rottke}}, \bibinfo {author} {\bibfnamefont
  {C.}~\bibnamefont {Trump}}, \bibinfo {author} {\bibfnamefont
  {M.}~\bibnamefont {Wittmann}}, \bibinfo {author} {\bibfnamefont
  {G.}~\bibnamefont {Korn}},  \emph {et~al.},\ }\href@noop {} {\bibfield
  {journal} {\bibinfo  {journal} {Physical review letters}\ }\textbf {\bibinfo
  {volume} {84}},\ \bibinfo {pages} {447} (\bibinfo {year} {2000})}\BibitemShut
  {NoStop}%
\bibitem [{\citenamefont {Rudenko}\ \emph {et~al.}(2008)\citenamefont
  {Rudenko}, \citenamefont {Ergler}, \citenamefont {Zrost}, \citenamefont
  {Feuerstein}, \citenamefont {De~Jesus}, \citenamefont {Schr{\"o}ter},
  \citenamefont {Moshammer},\ and\ \citenamefont {Ullrich}}]{rudenko_jpb08}%
  \BibitemOpen
  \bibfield  {author} {\bibinfo {author} {\bibfnamefont {A.}~\bibnamefont
  {Rudenko}}, \bibinfo {author} {\bibfnamefont {T.}~\bibnamefont {Ergler}},
  \bibinfo {author} {\bibfnamefont {K.}~\bibnamefont {Zrost}}, \bibinfo
  {author} {\bibfnamefont {B.}~\bibnamefont {Feuerstein}}, \bibinfo {author}
  {\bibfnamefont {V.}~\bibnamefont {De~Jesus}}, \bibinfo {author}
  {\bibfnamefont {C.}~\bibnamefont {Schr{\"o}ter}}, \bibinfo {author}
  {\bibfnamefont {R.}~\bibnamefont {Moshammer}}, \ and\ \bibinfo {author}
  {\bibfnamefont {J.}~\bibnamefont {Ullrich}},\ }\href@noop {} {\bibfield
  {journal} {\bibinfo  {journal} {Journal of Physics B: Atomic, Molecular and
  Optical Physics}\ }\textbf {\bibinfo {volume} {41}},\ \bibinfo {pages}
  {081006} (\bibinfo {year} {2008})}\BibitemShut {NoStop}%
\bibitem [{\citenamefont {Becker}\ \emph {et~al.}(2012)\citenamefont {Becker},
  \citenamefont {Liu}, \citenamefont {Ho},\ and\ \citenamefont
  {Eberly}}]{becker-rmp12}%
  \BibitemOpen
  \bibfield  {author} {\bibinfo {author} {\bibfnamefont {W.}~\bibnamefont
  {Becker}}, \bibinfo {author} {\bibfnamefont {X.}~\bibnamefont {Liu}},
  \bibinfo {author} {\bibfnamefont {P.~J.}\ \bibnamefont {Ho}}, \ and\ \bibinfo
  {author} {\bibfnamefont {J.~H.}\ \bibnamefont {Eberly}},\ }\href@noop {}
  {\bibfield  {journal} {\bibinfo  {journal} {Rev. Mod. Phys.}\ }\textbf
  {\bibinfo {volume} {84}},\ \bibinfo {pages} {1011} (\bibinfo {year}
  {2012})}\BibitemShut {NoStop}%
\bibitem [{\citenamefont {Larochelle}\ \emph {et~al.}(1998)\citenamefont
  {Larochelle}, \citenamefont {Talebpour},\ and\ \citenamefont
  {Chin}}]{chin_jpb98}%
  \BibitemOpen
  \bibfield  {author} {\bibinfo {author} {\bibfnamefont {S.}~\bibnamefont
  {Larochelle}}, \bibinfo {author} {\bibfnamefont {A.}~\bibnamefont
  {Talebpour}}, \ and\ \bibinfo {author} {\bibfnamefont {S.-L.}\ \bibnamefont
  {Chin}},\ }\href@noop {} {\bibfield  {journal} {\bibinfo  {journal} {Journal
  of Physics B: Atomic, Molecular and Optical Physics}\ }\textbf {\bibinfo
  {volume} {31}},\ \bibinfo {pages} {1201} (\bibinfo {year}
  {1998})}\BibitemShut {NoStop}%
\bibitem [{\citenamefont {Kuchiev}(1987)}]{kuchiev-jetpl87}%
  \BibitemOpen
  \bibfield  {author} {\bibinfo {author} {\bibfnamefont {M.~Y.}\ \bibnamefont
  {Kuchiev}},\ }\href@noop {} {\bibfield  {journal} {\bibinfo  {journal} {JETP
  Lett}\ }\textbf {\bibinfo {volume} {45}},\ \bibinfo {pages} {404} (\bibinfo
  {year} {1987})}\BibitemShut {NoStop}%
\bibitem [{\citenamefont {Corkum}(1993)}]{corkum-prl93}%
  \BibitemOpen
  \bibfield  {author} {\bibinfo {author} {\bibfnamefont {P.~B.}\ \bibnamefont
  {Corkum}},\ }\href@noop {} {\bibfield  {journal} {\bibinfo  {journal} {Phys.
  Rev. Lett.}\ }\textbf {\bibinfo {volume} {71}},\ \bibinfo {pages} {1994}
  (\bibinfo {year} {1993})}\BibitemShut {NoStop}%
\bibitem [{\citenamefont {Kopold}\ \emph {et~al.}(2000)\citenamefont {Kopold},
  \citenamefont {Becker}, \citenamefont {Rottke},\ and\ \citenamefont
  {Sandner}}]{kopold_prl00}%
  \BibitemOpen
  \bibfield  {author} {\bibinfo {author} {\bibfnamefont {R.}~\bibnamefont
  {Kopold}}, \bibinfo {author} {\bibfnamefont {W.}~\bibnamefont {Becker}},
  \bibinfo {author} {\bibfnamefont {H.}~\bibnamefont {Rottke}}, \ and\ \bibinfo
  {author} {\bibfnamefont {W.}~\bibnamefont {Sandner}},\ }\href@noop {}
  {\bibfield  {journal} {\bibinfo  {journal} {Physical review letters}\
  }\textbf {\bibinfo {volume} {85}},\ \bibinfo {pages} {3781} (\bibinfo {year}
  {2000})}\BibitemShut {NoStop}%
\bibitem [{\citenamefont {Agostini}(2024)}]{agostini_nobel24}%
  \BibitemOpen
  \bibfield  {author} {\bibinfo {author} {\bibfnamefont {P.}~\bibnamefont
  {Agostini}},\ }\href {\doibase 10.1103/RevModPhys.96.030501} {\bibfield
  {journal} {\bibinfo  {journal} {Rev. Mod. Phys.}\ }\textbf {\bibinfo {volume}
  {96}},\ \bibinfo {pages} {030501} (\bibinfo {year} {2024})}\BibitemShut
  {NoStop}%
\bibitem [{\citenamefont {L'Huillier}(2024)}]{huillier_nobel24}%
  \BibitemOpen
  \bibfield  {author} {\bibinfo {author} {\bibfnamefont {A.}~\bibnamefont
  {L'Huillier}},\ }\href {\doibase 10.1103/RevModPhys.96.030503} {\bibfield
  {journal} {\bibinfo  {journal} {Rev. Mod. Phys.}\ }\textbf {\bibinfo {volume}
  {96}},\ \bibinfo {pages} {030503} (\bibinfo {year} {2024})}\BibitemShut
  {NoStop}%
\bibitem [{\citenamefont {Krausz}(2024)}]{krausz_nobel24}%
  \BibitemOpen
  \bibfield  {author} {\bibinfo {author} {\bibfnamefont {F.}~\bibnamefont
  {Krausz}},\ }\href {\doibase 10.1103/RevModPhys.96.030502} {\bibfield
  {journal} {\bibinfo  {journal} {Rev. Mod. Phys.}\ }\textbf {\bibinfo {volume}
  {96}},\ \bibinfo {pages} {030502} (\bibinfo {year} {2024})}\BibitemShut
  {NoStop}%
\bibitem [{\citenamefont {Dimauro}\ and\ \citenamefont
  {Agostini}(1995)}]{agostini-rev95}%
  \BibitemOpen
  \bibfield  {author} {\bibinfo {author} {\bibfnamefont {L.}~\bibnamefont
  {Dimauro}}\ and\ \bibinfo {author} {\bibfnamefont {P.}~\bibnamefont
  {Agostini}},\ }in\ \href@noop {} {\emph {\bibinfo {booktitle} {Advances in
  Atomic, Molecular, and Optical Physics}}},\ Vol.~\bibinfo {volume} {35}\
  (\bibinfo  {publisher} {Elsevier},\ \bibinfo {year} {1995})\ pp.\ \bibinfo
  {pages} {79--120}\BibitemShut {NoStop}%
\bibitem [{\citenamefont {Zon}(1999)}]{zon-jetp99}%
  \BibitemOpen
  \bibfield  {author} {\bibinfo {author} {\bibfnamefont {B.}~\bibnamefont
  {Zon}},\ }\href@noop {} {\bibfield  {journal} {\bibinfo  {journal} {Journal
  of Experimental and Theoretical Physics}\ }\textbf {\bibinfo {volume} {89}},\
  \bibinfo {pages} {219} (\bibinfo {year} {1999})}\BibitemShut {NoStop}%
\bibitem [{\citenamefont {Eichmann}\ \emph {et~al.}(2000)\citenamefont
  {Eichmann}, \citenamefont {D\"orr}, \citenamefont {Maeda}, \citenamefont
  {Becker},\ and\ \citenamefont {Sandner}}]{becker-prl00}%
  \BibitemOpen
  \bibfield  {author} {\bibinfo {author} {\bibfnamefont {U.}~\bibnamefont
  {Eichmann}}, \bibinfo {author} {\bibfnamefont {M.}~\bibnamefont {D\"orr}},
  \bibinfo {author} {\bibfnamefont {H.}~\bibnamefont {Maeda}}, \bibinfo
  {author} {\bibfnamefont {W.}~\bibnamefont {Becker}}, \ and\ \bibinfo {author}
  {\bibfnamefont {W.}~\bibnamefont {Sandner}},\ }\href {\doibase
  10.1103/PhysRevLett.84.3550} {\bibfield  {journal} {\bibinfo  {journal}
  {Phys. Rev. Lett.}\ }\textbf {\bibinfo {volume} {84}},\ \bibinfo {pages}
  {3550} (\bibinfo {year} {2000})}\BibitemShut {NoStop}%
\bibitem [{\citenamefont {Popruzhenko}\ and\ \citenamefont
  {Tyurin}(2023)}]{tyurin-leb23}%
  \BibitemOpen
  \bibfield  {author} {\bibinfo {author} {\bibfnamefont {S.}~\bibnamefont
  {Popruzhenko}}\ and\ \bibinfo {author} {\bibfnamefont {D.}~\bibnamefont
  {Tyurin}},\ }\href@noop {} {\bibfield  {journal} {\bibinfo  {journal}
  {Bulletin of the Lebedev Physics Institute}\ }\textbf {\bibinfo {volume}
  {50}},\ \bibinfo {pages} {S922} (\bibinfo {year} {2023})}\BibitemShut
  {NoStop}%
\bibitem [{\citenamefont {Lein}\ \emph {et~al.}(2000)\citenamefont {Lein},
  \citenamefont {Gross},\ and\ \citenamefont {Engel}}]{lein_prl00}%
  \BibitemOpen
  \bibfield  {author} {\bibinfo {author} {\bibfnamefont {M.}~\bibnamefont
  {Lein}}, \bibinfo {author} {\bibfnamefont {E.~K.}\ \bibnamefont {Gross}}, \
  and\ \bibinfo {author} {\bibfnamefont {V.}~\bibnamefont {Engel}},\
  }\href@noop {} {\bibfield  {journal} {\bibinfo  {journal} {Phys. Rev. Lett.}\
  }\textbf {\bibinfo {volume} {85}},\ \bibinfo {pages} {4707} (\bibinfo {year}
  {2000})}\BibitemShut {NoStop}%
\bibitem [{\citenamefont {Klaiber}\ \emph {et~al.}(2007)\citenamefont
  {Klaiber}, \citenamefont {Hatsagortsyan},\ and\ \citenamefont
  {Keitel}}]{keitel-pra07}%
  \BibitemOpen
  \bibfield  {author} {\bibinfo {author} {\bibfnamefont {M.}~\bibnamefont
  {Klaiber}}, \bibinfo {author} {\bibfnamefont {K.~Z.}\ \bibnamefont
  {Hatsagortsyan}}, \ and\ \bibinfo {author} {\bibfnamefont {C.~H.}\
  \bibnamefont {Keitel}},\ }\href@noop {} {\bibfield  {journal} {\bibinfo
  {journal} {Physical Review A—Atomic, Molecular, and Optical Physics}\
  }\textbf {\bibinfo {volume} {75}},\ \bibinfo {pages} {063413} (\bibinfo
  {year} {2007})}\BibitemShut {NoStop}%
\bibitem [{\citenamefont {Shvetsov-Shilovski}\ \emph
  {et~al.}(2008)\citenamefont {Shvetsov-Shilovski}, \citenamefont
  {Goreslavski}, \citenamefont {Popruzhenko},\ and\ \citenamefont
  {Becker}}]{kolya-pra08}%
  \BibitemOpen
  \bibfield  {author} {\bibinfo {author} {\bibfnamefont {N.}~\bibnamefont
  {Shvetsov-Shilovski}}, \bibinfo {author} {\bibfnamefont {S.}~\bibnamefont
  {Goreslavski}}, \bibinfo {author} {\bibfnamefont {S.}~\bibnamefont
  {Popruzhenko}}, \ and\ \bibinfo {author} {\bibfnamefont {W.}~\bibnamefont
  {Becker}},\ }\href@noop {} {\bibfield  {journal} {\bibinfo  {journal}
  {Physical Review A—Atomic, Molecular, and Optical Physics}\ }\textbf
  {\bibinfo {volume} {77}},\ \bibinfo {pages} {063405} (\bibinfo {year}
  {2008})}\BibitemShut {NoStop}%
\bibitem [{\citenamefont {Liu}\ and\ \citenamefont {Figueira~de
  Morisson~Faria}(2004)}]{carla_prl04}%
  \BibitemOpen
  \bibfield  {author} {\bibinfo {author} {\bibfnamefont {X.}~\bibnamefont
  {Liu}}\ and\ \bibinfo {author} {\bibfnamefont {C.}~\bibnamefont {Figueira~de
  Morisson~Faria}},\ }\href@noop {} {\bibfield  {journal} {\bibinfo  {journal}
  {Physical review letters}\ }\textbf {\bibinfo {volume} {92}},\ \bibinfo
  {pages} {133006} (\bibinfo {year} {2004})}\BibitemShut {NoStop}%
\bibitem [{\citenamefont {Milo{\v{s}}evi{\'c}}\ \emph
  {et~al.}(2006)\citenamefont {Milo{\v{s}}evi{\'c}}, \citenamefont {Paulus},
  \citenamefont {Bauer},\ and\ \citenamefont {Becker}}]{paulus_jpb06}%
  \BibitemOpen
  \bibfield  {author} {\bibinfo {author} {\bibfnamefont {D.}~\bibnamefont
  {Milo{\v{s}}evi{\'c}}}, \bibinfo {author} {\bibfnamefont {G.}~\bibnamefont
  {Paulus}}, \bibinfo {author} {\bibfnamefont {D.}~\bibnamefont {Bauer}}, \
  and\ \bibinfo {author} {\bibfnamefont {W.}~\bibnamefont {Becker}},\
  }\href@noop {} {\bibfield  {journal} {\bibinfo  {journal} {Journal of Physics
  B: Atomic, Molecular and Optical Physics}\ }\textbf {\bibinfo {volume}
  {39}},\ \bibinfo {pages} {R203} (\bibinfo {year} {2006})}\BibitemShut
  {NoStop}%
\bibitem [{\citenamefont {Bergues}\ \emph
  {et~al.}(2012{\natexlab{a}})\citenamefont {Bergues}, \citenamefont
  {K{\"u}bel}, \citenamefont {Johnson}, \citenamefont {Fischer}, \citenamefont
  {Camus}, \citenamefont {Betsch}, \citenamefont {Herrwerth}, \citenamefont
  {Senftleben}, \citenamefont {Sayler}, \citenamefont {Rathje} \emph
  {et~al.}}]{bergues_nc12}%
  \BibitemOpen
  \bibfield  {author} {\bibinfo {author} {\bibfnamefont {B.}~\bibnamefont
  {Bergues}}, \bibinfo {author} {\bibfnamefont {M.}~\bibnamefont {K{\"u}bel}},
  \bibinfo {author} {\bibfnamefont {N.~G.}\ \bibnamefont {Johnson}}, \bibinfo
  {author} {\bibfnamefont {B.}~\bibnamefont {Fischer}}, \bibinfo {author}
  {\bibfnamefont {N.}~\bibnamefont {Camus}}, \bibinfo {author} {\bibfnamefont
  {K.~J.}\ \bibnamefont {Betsch}}, \bibinfo {author} {\bibfnamefont
  {O.}~\bibnamefont {Herrwerth}}, \bibinfo {author} {\bibfnamefont
  {A.}~\bibnamefont {Senftleben}}, \bibinfo {author} {\bibfnamefont {A.~M.}\
  \bibnamefont {Sayler}}, \bibinfo {author} {\bibfnamefont {T.}~\bibnamefont
  {Rathje}},  \emph {et~al.},\ }\href@noop {} {\bibfield  {journal} {\bibinfo
  {journal} {Nature communications}\ }\textbf {\bibinfo {volume} {3}},\
  \bibinfo {pages} {813} (\bibinfo {year} {2012}{\natexlab{a}})}\BibitemShut
  {NoStop}%
\bibitem [{\citenamefont {K{\"u}bel}\ \emph {et~al.}(2013)\citenamefont
  {K{\"u}bel}, \citenamefont {Kling}, \citenamefont {Betsch}, \citenamefont
  {Camus}, \citenamefont {Kaldun}, \citenamefont {Kleineberg}, \citenamefont
  {Ben-Itzhak}, \citenamefont {Jones}, \citenamefont {Paulus}, \citenamefont
  {Pfeifer} \emph {et~al.}}]{kubel_pra13}%
  \BibitemOpen
  \bibfield  {author} {\bibinfo {author} {\bibfnamefont {M.}~\bibnamefont
  {K{\"u}bel}}, \bibinfo {author} {\bibfnamefont {N.~G.}\ \bibnamefont
  {Kling}}, \bibinfo {author} {\bibfnamefont {K.~J.}\ \bibnamefont {Betsch}},
  \bibinfo {author} {\bibfnamefont {N.}~\bibnamefont {Camus}}, \bibinfo
  {author} {\bibfnamefont {A.}~\bibnamefont {Kaldun}}, \bibinfo {author}
  {\bibfnamefont {U.}~\bibnamefont {Kleineberg}}, \bibinfo {author}
  {\bibfnamefont {I.}~\bibnamefont {Ben-Itzhak}}, \bibinfo {author}
  {\bibfnamefont {R.}~\bibnamefont {Jones}}, \bibinfo {author} {\bibfnamefont
  {G.}~\bibnamefont {Paulus}}, \bibinfo {author} {\bibfnamefont
  {T.}~\bibnamefont {Pfeifer}},  \emph {et~al.},\ }\href@noop {} {\bibfield
  {journal} {\bibinfo  {journal} {Physical Review A—Atomic, Molecular, and
  Optical Physics}\ }\textbf {\bibinfo {volume} {88}},\ \bibinfo {pages}
  {023418} (\bibinfo {year} {2013})}\BibitemShut {NoStop}%
\bibitem [{\citenamefont {Ben}\ \emph {et~al.}(2016)\citenamefont {Ben},
  \citenamefont {Wang}, \citenamefont {Xu}, \citenamefont {Guo},\ and\
  \citenamefont {Liu}}]{ben_oe16}%
  \BibitemOpen
  \bibfield  {author} {\bibinfo {author} {\bibfnamefont {S.}~\bibnamefont
  {Ben}}, \bibinfo {author} {\bibfnamefont {T.}~\bibnamefont {Wang}}, \bibinfo
  {author} {\bibfnamefont {T.}~\bibnamefont {Xu}}, \bibinfo {author}
  {\bibfnamefont {J.}~\bibnamefont {Guo}}, \ and\ \bibinfo {author}
  {\bibfnamefont {X.}~\bibnamefont {Liu}},\ }\href@noop {} {\bibfield
  {journal} {\bibinfo  {journal} {Optics Express}\ }\textbf {\bibinfo {volume}
  {24}},\ \bibinfo {pages} {7525} (\bibinfo {year} {2016})}\BibitemShut
  {NoStop}%
\bibitem [{\citenamefont {Becker}\ \emph {et~al.}(2018)\citenamefont {Becker},
  \citenamefont {Goreslavski}, \citenamefont {Milo{\v{s}}evi{\'c}},\ and\
  \citenamefont {Paulus}}]{paulus_jpb18}%
  \BibitemOpen
  \bibfield  {author} {\bibinfo {author} {\bibfnamefont {W.}~\bibnamefont
  {Becker}}, \bibinfo {author} {\bibfnamefont {S.}~\bibnamefont {Goreslavski}},
  \bibinfo {author} {\bibfnamefont {D.}~\bibnamefont {Milo{\v{s}}evi{\'c}}}, \
  and\ \bibinfo {author} {\bibfnamefont {G.}~\bibnamefont {Paulus}},\
  }\href@noop {} {\bibfield  {journal} {\bibinfo  {journal} {Journal of Physics
  B: Atomic, Molecular and Optical Physics}\ }\textbf {\bibinfo {volume}
  {51}},\ \bibinfo {pages} {162002} (\bibinfo {year} {2018})}\BibitemShut
  {NoStop}%
\bibitem [{\citenamefont {Li}\ \emph {et~al.}(2020)\citenamefont {Li},
  \citenamefont {Sierra-Costa}, \citenamefont {Michie}, \citenamefont
  {Ben-Itzhak},\ and\ \citenamefont {Dantus}}]{itzhak_nc20}%
  \BibitemOpen
  \bibfield  {author} {\bibinfo {author} {\bibfnamefont {S.}~\bibnamefont
  {Li}}, \bibinfo {author} {\bibfnamefont {D.}~\bibnamefont {Sierra-Costa}},
  \bibinfo {author} {\bibfnamefont {M.~J.}\ \bibnamefont {Michie}}, \bibinfo
  {author} {\bibfnamefont {I.}~\bibnamefont {Ben-Itzhak}}, \ and\ \bibinfo
  {author} {\bibfnamefont {M.}~\bibnamefont {Dantus}},\ }\href@noop {}
  {\bibfield  {journal} {\bibinfo  {journal} {Communications Physics}\ }\textbf
  {\bibinfo {volume} {3}},\ \bibinfo {pages} {35} (\bibinfo {year}
  {2020})}\BibitemShut {NoStop}%
\bibitem [{\citenamefont {Oppenheimer}(1928)}]{opp_pr28}%
  \BibitemOpen
  \bibfield  {author} {\bibinfo {author} {\bibfnamefont {J.~R.}\ \bibnamefont
  {Oppenheimer}},\ }\href {\doibase 10.1103/PhysRev.31.66} {\bibfield
  {journal} {\bibinfo  {journal} {Phys. Rev.}\ }\textbf {\bibinfo {volume}
  {31}},\ \bibinfo {pages} {66} (\bibinfo {year} {1928})}\BibitemShut {NoStop}%
\bibitem [{\citenamefont {Keldysh}(1965)}]{keldysh-jetp65}%
  \BibitemOpen
  \bibfield  {author} {\bibinfo {author} {\bibfnamefont {L.}~\bibnamefont
  {Keldysh}},\ }\href@noop {} {\bibfield  {journal} {\bibinfo  {journal}
  {Soviet Phys. JETP}\ }\textbf {\bibinfo {volume} {20}} (\bibinfo {year}
  {1965})}\BibitemShut {NoStop}%
\bibitem [{\citenamefont {Perelomov}\ \emph {et~al.}(1966)\citenamefont
  {Perelomov}, \citenamefont {Popov},\ and\ \citenamefont
  {Terent’ev}}]{popov_jetp66}%
  \BibitemOpen
  \bibfield  {author} {\bibinfo {author} {\bibfnamefont {A.}~\bibnamefont
  {Perelomov}}, \bibinfo {author} {\bibfnamefont {V.}~\bibnamefont {Popov}}, \
  and\ \bibinfo {author} {\bibfnamefont {M.}~\bibnamefont {Terent’ev}},\
  }\href@noop {} {\bibfield  {journal} {\bibinfo  {journal} {Zh. Eksp. Teor.
  Fiz}\ }\textbf {\bibinfo {volume} {50}},\ \bibinfo {pages} {1393} (\bibinfo
  {year} {1966})}\BibitemShut {NoStop}%
\bibitem [{\citenamefont {Zeldovich}(1960)}]{zeldovich_jetp60}%
  \BibitemOpen
  \bibfield  {author} {\bibinfo {author} {\bibfnamefont {Y.~B.}\ \bibnamefont
  {Zeldovich}},\ }\href@noop {} {\bibfield  {journal} {\bibinfo  {journal}
  {Sov. Phys. JETP}\ }\textbf {\bibinfo {volume} {38}},\ \bibinfo {pages}
  {1123} (\bibinfo {year} {1960})}\BibitemShut {NoStop}%
\bibitem [{\citenamefont {Galitsky}\ and\ \citenamefont
  {Cheltsov}(1964)}]{galitsky_np64}%
  \BibitemOpen
  \bibfield  {author} {\bibinfo {author} {\bibfnamefont {V.}~\bibnamefont
  {Galitsky}}\ and\ \bibinfo {author} {\bibfnamefont {V.}~\bibnamefont
  {Cheltsov}},\ }\href@noop {} {\bibfield  {journal} {\bibinfo  {journal}
  {Nuclear Physics}\ }\textbf {\bibinfo {volume} {56}},\ \bibinfo {pages} {86}
  (\bibinfo {year} {1964})}\BibitemShut {NoStop}%
\bibitem [{\citenamefont {Blank}\ and\ \citenamefont
  {P{\l}oszajczak}(2008)}]{blank_rpp08}%
  \BibitemOpen
  \bibfield  {author} {\bibinfo {author} {\bibfnamefont {B.}~\bibnamefont
  {Blank}}\ and\ \bibinfo {author} {\bibfnamefont {M.}~\bibnamefont
  {P{\l}oszajczak}},\ }\href@noop {} {\bibfield  {journal} {\bibinfo  {journal}
  {Reports on Progress in Physics}\ }\textbf {\bibinfo {volume} {71}},\
  \bibinfo {pages} {046301} (\bibinfo {year} {2008})}\BibitemShut {NoStop}%
\bibitem [{\citenamefont {Popov}(2004)}]{popov-usp04}%
  \BibitemOpen
  \bibfield  {author} {\bibinfo {author} {\bibfnamefont {V.~S.}\ \bibnamefont
  {Popov}},\ }\href@noop {} {\bibfield  {journal} {\bibinfo  {journal}
  {Physics-Uspekhi}\ }\textbf {\bibinfo {volume} {47}},\ \bibinfo {pages} {855}
  (\bibinfo {year} {2004})}\BibitemShut {NoStop}%
\bibitem [{\citenamefont {Popov}(2005)}]{popov2005}%
  \BibitemOpen
  \bibfield  {author} {\bibinfo {author} {\bibfnamefont {V.}~\bibnamefont
  {Popov}},\ }\href@noop {} {\bibfield  {journal} {\bibinfo  {journal} {Physics
  of Atomic Nuclei}\ }\textbf {\bibinfo {volume} {68}},\ \bibinfo {pages} {686}
  (\bibinfo {year} {2005})}\BibitemShut {NoStop}%
\bibitem [{\citenamefont {Salieres}\ \emph {et~al.}(2001)\citenamefont
  {Salieres}, \citenamefont {Carr{\'e}}, \citenamefont {Le~D{\'e}roff},
  \citenamefont {Grasbon}, \citenamefont {Paulus}, \citenamefont {Walther},
  \citenamefont {Kopold}, \citenamefont {Becker}, \citenamefont {Milosevic},
  \citenamefont {Sanpera} \emph {et~al.}}]{salieres_sci01}%
  \BibitemOpen
  \bibfield  {author} {\bibinfo {author} {\bibfnamefont {P.}~\bibnamefont
  {Salieres}}, \bibinfo {author} {\bibfnamefont {B.}~\bibnamefont {Carr{\'e}}},
  \bibinfo {author} {\bibfnamefont {L.}~\bibnamefont {Le~D{\'e}roff}}, \bibinfo
  {author} {\bibfnamefont {F.}~\bibnamefont {Grasbon}}, \bibinfo {author}
  {\bibfnamefont {G.}~\bibnamefont {Paulus}}, \bibinfo {author} {\bibfnamefont
  {H.}~\bibnamefont {Walther}}, \bibinfo {author} {\bibfnamefont
  {R.}~\bibnamefont {Kopold}}, \bibinfo {author} {\bibfnamefont
  {W.}~\bibnamefont {Becker}}, \bibinfo {author} {\bibfnamefont
  {D.}~\bibnamefont {Milosevic}}, \bibinfo {author} {\bibfnamefont
  {A.}~\bibnamefont {Sanpera}},  \emph {et~al.},\ }\href@noop {} {\bibfield
  {journal} {\bibinfo  {journal} {Science}\ }\textbf {\bibinfo {volume}
  {292}},\ \bibinfo {pages} {902} (\bibinfo {year} {2001})}\BibitemShut
  {NoStop}%
\bibitem [{\citenamefont {Eckhardt}\ and\ \citenamefont
  {Sacha}(2001)}]{eckhardt_epj01}%
  \BibitemOpen
  \bibfield  {author} {\bibinfo {author} {\bibfnamefont {B.}~\bibnamefont
  {Eckhardt}}\ and\ \bibinfo {author} {\bibfnamefont {K.}~\bibnamefont
  {Sacha}},\ }\href@noop {} {\bibfield  {journal} {\bibinfo  {journal} {EPL
  (Europhysics Letters)}\ }\textbf {\bibinfo {volume} {56}},\ \bibinfo {pages}
  {651} (\bibinfo {year} {2001})}\BibitemShut {NoStop}%
\bibitem [{\citenamefont {Sacha}\ and\ \citenamefont
  {Eckhardt}(2001)}]{eckhardt_pra01}%
  \BibitemOpen
  \bibfield  {author} {\bibinfo {author} {\bibfnamefont {K.}~\bibnamefont
  {Sacha}}\ and\ \bibinfo {author} {\bibfnamefont {B.}~\bibnamefont
  {Eckhardt}},\ }\href@noop {} {\bibfield  {journal} {\bibinfo  {journal}
  {Physical Review A}\ }\textbf {\bibinfo {volume} {63}},\ \bibinfo {pages}
  {043414} (\bibinfo {year} {2001})}\BibitemShut {NoStop}%
\bibitem [{\citenamefont {Eckhardt}\ and\ \citenamefont
  {Sacha}(2006)}]{eckhardt_jpb06}%
  \BibitemOpen
  \bibfield  {author} {\bibinfo {author} {\bibfnamefont {B.}~\bibnamefont
  {Eckhardt}}\ and\ \bibinfo {author} {\bibfnamefont {K.}~\bibnamefont
  {Sacha}},\ }\href@noop {} {\bibfield  {journal} {\bibinfo  {journal} {Journal
  of Physics B: Atomic, Molecular and Optical Physics}\ }\textbf {\bibinfo
  {volume} {39}},\ \bibinfo {pages} {3865} (\bibinfo {year}
  {2006})}\BibitemShut {NoStop}%
\bibitem [{\citenamefont {Hall}\ \emph {et~al.}(2009)\citenamefont {Hall},
  \citenamefont {Saad}, \citenamefont {Sen},\ and\ \citenamefont
  {Ciftci}}]{hall2009energies}%
  \BibitemOpen
  \bibfield  {author} {\bibinfo {author} {\bibfnamefont {R.~L.}\ \bibnamefont
  {Hall}}, \bibinfo {author} {\bibfnamefont {N.}~\bibnamefont {Saad}}, \bibinfo
  {author} {\bibfnamefont {K.}~\bibnamefont {Sen}}, \ and\ \bibinfo {author}
  {\bibfnamefont {H.}~\bibnamefont {Ciftci}},\ }\href@noop {} {\bibfield
  {journal} {\bibinfo  {journal} {Physical Review A—Atomic, Molecular, and
  Optical Physics}\ }\textbf {\bibinfo {volume} {80}},\ \bibinfo {pages}
  {032507} (\bibinfo {year} {2009})}\BibitemShut {NoStop}%
\bibitem [{\citenamefont {Fern{\'a}ndez}(1991)}]{fernandez1991analytical}%
  \BibitemOpen
  \bibfield  {author} {\bibinfo {author} {\bibfnamefont {F.}~\bibnamefont
  {Fern{\'a}ndez}},\ }\href@noop {} {\bibfield  {journal} {\bibinfo  {journal}
  {Journal of Physics A: Mathematical and General}\ }\textbf {\bibinfo {volume}
  {24}},\ \bibinfo {pages} {1351} (\bibinfo {year} {1991})}\BibitemShut
  {NoStop}%
\bibitem [{\citenamefont {Lima}\ and\ \citenamefont
  {Miranda}(1981)}]{lima1981atoms}%
  \BibitemOpen
  \bibfield  {author} {\bibinfo {author} {\bibfnamefont {C.}~\bibnamefont
  {Lima}}\ and\ \bibinfo {author} {\bibfnamefont {L.}~\bibnamefont {Miranda}},\
  }\href@noop {} {\bibfield  {journal} {\bibinfo  {journal} {Physical Review
  A}\ }\textbf {\bibinfo {volume} {23}},\ \bibinfo {pages} {3335} (\bibinfo
  {year} {1981})}\BibitemShut {NoStop}%
\bibitem [{\citenamefont {Silaev}\ \emph {et~al.}(2010)\citenamefont {Silaev},
  \citenamefont {Ryabikin},\ and\ \citenamefont
  {Vvedenskii}}]{silaev2010strong}%
  \BibitemOpen
  \bibfield  {author} {\bibinfo {author} {\bibfnamefont {A.}~\bibnamefont
  {Silaev}}, \bibinfo {author} {\bibfnamefont {M.~Y.}\ \bibnamefont
  {Ryabikin}}, \ and\ \bibinfo {author} {\bibfnamefont {N.}~\bibnamefont
  {Vvedenskii}},\ }\href@noop {} {\bibfield  {journal} {\bibinfo  {journal}
  {Physical Review A—Atomic, Molecular, and Optical Physics}\ }\textbf
  {\bibinfo {volume} {82}},\ \bibinfo {pages} {033416} (\bibinfo {year}
  {2010})}\BibitemShut {NoStop}%
\bibitem [{\citenamefont {Gebremedhin}\ and\ \citenamefont
  {Weatherford}(2014)}]{gebremedhin2014calculations}%
  \BibitemOpen
  \bibfield  {author} {\bibinfo {author} {\bibfnamefont {D.~H.}\ \bibnamefont
  {Gebremedhin}}\ and\ \bibinfo {author} {\bibfnamefont {C.~A.}\ \bibnamefont
  {Weatherford}},\ }\href@noop {} {\bibfield  {journal} {\bibinfo  {journal}
  {Physical Review E}\ }\textbf {\bibinfo {volume} {89}},\ \bibinfo {pages}
  {053319} (\bibinfo {year} {2014})}\BibitemShut {NoStop}%
\bibitem [{\citenamefont {Bader}\ \emph {et~al.}(2013)\citenamefont {Bader},
  \citenamefont {Blanes},\ and\ \citenamefont {Casas}}]{bader2013}%
  \BibitemOpen
  \bibfield  {author} {\bibinfo {author} {\bibfnamefont {P.}~\bibnamefont
  {Bader}}, \bibinfo {author} {\bibfnamefont {S.}~\bibnamefont {Blanes}}, \
  and\ \bibinfo {author} {\bibfnamefont {F.}~\bibnamefont {Casas}},\
  }\href@noop {} {\bibfield  {journal} {\bibinfo  {journal} {The Journal of
  chemical physics}\ }\textbf {\bibinfo {volume} {139}} (\bibinfo {year}
  {2013})}\BibitemShut {NoStop}%
\bibitem [{\citenamefont {Weideman}\ and\ \citenamefont
  {Herbst}(1986)}]{weideman1986split}%
  \BibitemOpen
  \bibfield  {author} {\bibinfo {author} {\bibfnamefont {J.~A.~C.}\
  \bibnamefont {Weideman}}\ and\ \bibinfo {author} {\bibfnamefont {B.~M.}\
  \bibnamefont {Herbst}},\ }\href@noop {} {\bibfield  {journal} {\bibinfo
  {journal} {SIAM Journal on Numerical Analysis}\ }\textbf {\bibinfo {volume}
  {23}},\ \bibinfo {pages} {485} (\bibinfo {year} {1986})}\BibitemShut
  {NoStop}%
\bibitem [{\citenamefont {Neuhasuer}\ and\ \citenamefont
  {Baer}(1989)}]{neuhasuer1989}%
  \BibitemOpen
  \bibfield  {author} {\bibinfo {author} {\bibfnamefont {D.}~\bibnamefont
  {Neuhasuer}}\ and\ \bibinfo {author} {\bibfnamefont {M.}~\bibnamefont
  {Baer}},\ }\href@noop {} {\bibfield  {journal} {\bibinfo  {journal} {The
  Journal of chemical physics}\ }\textbf {\bibinfo {volume} {90}},\ \bibinfo
  {pages} {4351} (\bibinfo {year} {1989})}\BibitemShut {NoStop}%
\bibitem [{\citenamefont {Smirnov}\ and\ \citenamefont
  {Chibisov}(1966)}]{smirnov_jetp66}%
  \BibitemOpen
  \bibfield  {author} {\bibinfo {author} {\bibfnamefont {B.}~\bibnamefont
  {Smirnov}}\ and\ \bibinfo {author} {\bibfnamefont {M.}~\bibnamefont
  {Chibisov}},\ }\href@noop {} {\bibfield  {journal} {\bibinfo  {journal} {Sov.
  Phys. JETP}\ }\textbf {\bibinfo {volume} {22}},\ \bibinfo {pages} {23}
  (\bibinfo {year} {1966})}\BibitemShut {NoStop}%
\bibitem [{\citenamefont {Tong}\ and\ \citenamefont {Lin}(2005)}]{tong_jpb05}%
  \BibitemOpen
  \bibfield  {author} {\bibinfo {author} {\bibfnamefont {X.}~\bibnamefont
  {Tong}}\ and\ \bibinfo {author} {\bibfnamefont {C.}~\bibnamefont {Lin}},\
  }\href@noop {} {\bibfield  {journal} {\bibinfo  {journal} {Journal of Physics
  B: Atomic, Molecular and Optical Physics}\ }\textbf {\bibinfo {volume}
  {38}},\ \bibinfo {pages} {2593} (\bibinfo {year} {2005})}\BibitemShut
  {NoStop}%
\bibitem [{\citenamefont {Kostyukov}\ and\ \citenamefont
  {Golovanov}(2018)}]{kostyukov_pra18}%
  \BibitemOpen
  \bibfield  {author} {\bibinfo {author} {\bibfnamefont {I.~Y.}\ \bibnamefont
  {Kostyukov}}\ and\ \bibinfo {author} {\bibfnamefont {A.}~\bibnamefont
  {Golovanov}},\ }\href@noop {} {\bibfield  {journal} {\bibinfo  {journal}
  {Physical Review A}\ }\textbf {\bibinfo {volume} {98}},\ \bibinfo {pages}
  {043407} (\bibinfo {year} {2018})}\BibitemShut {NoStop}%
\bibitem [{\citenamefont {Popruzhenko}(2014)}]{poprz_jpb14}%
  \BibitemOpen
  \bibfield  {author} {\bibinfo {author} {\bibfnamefont {S.}~\bibnamefont
  {Popruzhenko}},\ }\href@noop {} {\bibfield  {journal} {\bibinfo  {journal}
  {Journal of Physics B: Atomic, Molecular and Optical Physics}\ }\textbf
  {\bibinfo {volume} {47}},\ \bibinfo {pages} {204001} (\bibinfo {year}
  {2014})}\BibitemShut {NoStop}%
\bibitem [{\citenamefont {Bauer}(2017)}]{bauer_book17}%
  \BibitemOpen
  \bibfield  {author} {\bibinfo {author} {\bibfnamefont {D.}~\bibnamefont
  {Bauer}},\ }\href@noop {} {\emph {\bibinfo {title} {Computational
  strong-field quantum dynamics}}}\ (\bibinfo  {publisher} {de Gruyter
  Berlin},\ \bibinfo {year} {2017})\BibitemShut {NoStop}%
\bibitem [{\citenamefont {Shvetsov-Shilovski}(2025)}]{sh-sh_lp25}%
  \BibitemOpen
  \bibfield  {author} {\bibinfo {author} {\bibfnamefont {N.}~\bibnamefont
  {Shvetsov-Shilovski}},\ }\href@noop {} {\bibfield  {journal} {\bibinfo
  {journal} {Laser Physics}\ }\textbf {\bibinfo {volume} {35}},\ \bibinfo
  {pages} {123001} (\bibinfo {year} {2025})}\BibitemShut {NoStop}%
\bibitem [{\citenamefont {Perelomov}\ \emph {et~al.}(1967)\citenamefont
  {Perelomov}, \citenamefont {Popov},\ and\ \citenamefont
  {Terent’ev}}]{perelomov_jetp67}%
  \BibitemOpen
  \bibfield  {author} {\bibinfo {author} {\bibfnamefont {A.}~\bibnamefont
  {Perelomov}}, \bibinfo {author} {\bibfnamefont {V.}~\bibnamefont {Popov}}, \
  and\ \bibinfo {author} {\bibfnamefont {M.}~\bibnamefont {Terent’ev}},\
  }\href@noop {} {\bibfield  {journal} {\bibinfo  {journal} {Zh. Eksp. Teor.
  Fiz}\ }\textbf {\bibinfo {volume} {52}},\ \bibinfo {pages} {514} (\bibinfo
  {year} {1967})}\BibitemShut {NoStop}%
\bibitem [{\citenamefont {Popruzhenko}\ \emph {et~al.}(2008)\citenamefont
  {Popruzhenko}, \citenamefont {Mur}, \citenamefont {Popov},\ and\
  \citenamefont {Bauer}}]{poprz_prl08}%
  \BibitemOpen
  \bibfield  {author} {\bibinfo {author} {\bibfnamefont {S.}~\bibnamefont
  {Popruzhenko}}, \bibinfo {author} {\bibfnamefont {V.}~\bibnamefont {Mur}},
  \bibinfo {author} {\bibfnamefont {V.}~\bibnamefont {Popov}}, \ and\ \bibinfo
  {author} {\bibfnamefont {D.}~\bibnamefont {Bauer}},\ }\href@noop {}
  {\bibfield  {journal} {\bibinfo  {journal} {Physical review letters}\
  }\textbf {\bibinfo {volume} {101}},\ \bibinfo {pages} {193003} (\bibinfo
  {year} {2008})}\BibitemShut {NoStop}%
\bibitem [{\citenamefont {Ankerhold}(2007)}]{ankerhold-07}%
  \BibitemOpen
  \bibfield  {author} {\bibinfo {author} {\bibfnamefont {J.}~\bibnamefont
  {Ankerhold}},\ }\href@noop {} {\emph {\bibinfo {title} {Quantum tunneling in
  complex systems: the semiclassical approach}}},\ Vol.\ \bibinfo {volume}
  {224}\ (\bibinfo  {publisher} {Springer},\ \bibinfo {year}
  {2007})\BibitemShut {NoStop}%
\bibitem [{\citenamefont {Razavy}(2013)}]{razavy_book13}%
  \BibitemOpen
  \bibfield  {author} {\bibinfo {author} {\bibfnamefont {M.}~\bibnamefont
  {Razavy}},\ }\href@noop {} {\emph {\bibinfo {title} {Quantum theory of
  tunneling}}}\ (\bibinfo  {publisher} {World Scientific},\ \bibinfo {year}
  {2013})\BibitemShut {NoStop}%
\bibitem [{\citenamefont {Mauger}\ \emph {et~al.}(2012)\citenamefont {Mauger},
  \citenamefont {Kamor}, \citenamefont {Chandre},\ and\ \citenamefont
  {Uzer}}]{mauger_prl12}%
  \BibitemOpen
  \bibfield  {author} {\bibinfo {author} {\bibfnamefont {F.}~\bibnamefont
  {Mauger}}, \bibinfo {author} {\bibfnamefont {A.}~\bibnamefont {Kamor}},
  \bibinfo {author} {\bibfnamefont {C.}~\bibnamefont {Chandre}}, \ and\
  \bibinfo {author} {\bibfnamefont {T.}~\bibnamefont {Uzer}},\ }\href@noop {}
  {\bibfield  {journal} {\bibinfo  {journal} {Physical Review Letters}\
  }\textbf {\bibinfo {volume} {108}},\ \bibinfo {pages} {063001} (\bibinfo
  {year} {2012})}\BibitemShut {NoStop}%
\bibitem [{\citenamefont {Tyurin}\ \emph {et~al.}(2025)\citenamefont {Tyurin},
  \citenamefont {Strelkov},\ and\ \citenamefont
  {Popruzhenko}}]{tyurin_arxiv25}%
  \BibitemOpen
  \bibfield  {author} {\bibinfo {author} {\bibfnamefont {D.}~\bibnamefont
  {Tyurin}}, \bibinfo {author} {\bibfnamefont {V.}~\bibnamefont {Strelkov}}, \
  and\ \bibinfo {author} {\bibfnamefont {S.}~\bibnamefont {Popruzhenko}},\
  }\href@noop {} {\bibfield  {journal} {\bibinfo  {journal} {arXiv preprint
  arXiv:2504.20583}\ } (\bibinfo {year} {2025})}\BibitemShut {NoStop}%
\bibitem [{\citenamefont {de~Morisson~Faria}\ and\ \citenamefont
  {Liu}(2011)}]{faria_jmo11}%
  \BibitemOpen
  \bibfield  {author} {\bibinfo {author} {\bibfnamefont {C.~F.}\ \bibnamefont
  {de~Morisson~Faria}}\ and\ \bibinfo {author} {\bibfnamefont {X.}~\bibnamefont
  {Liu}},\ }\href@noop {} {\bibfield  {journal} {\bibinfo  {journal} {Journal
  of Modern Optics}\ }\textbf {\bibinfo {volume} {58}},\ \bibinfo {pages}
  {1076} (\bibinfo {year} {2011})}\BibitemShut {NoStop}%
\bibitem [{\citenamefont {Bergues}\ \emph
  {et~al.}(2012{\natexlab{b}})\citenamefont {Bergues}, \citenamefont
  {K{\"u}bel}, \citenamefont {Johnson}, \citenamefont {Fischer}, \citenamefont
  {Camus}, \citenamefont {Betsch}, \citenamefont {Herrwerth}, \citenamefont
  {Senftleben}, \citenamefont {Sayler}, \citenamefont {Rathje} \emph
  {et~al.}}]{bergues_natcom12}%
  \BibitemOpen
  \bibfield  {author} {\bibinfo {author} {\bibfnamefont {B.}~\bibnamefont
  {Bergues}}, \bibinfo {author} {\bibfnamefont {M.}~\bibnamefont {K{\"u}bel}},
  \bibinfo {author} {\bibfnamefont {N.~G.}\ \bibnamefont {Johnson}}, \bibinfo
  {author} {\bibfnamefont {B.}~\bibnamefont {Fischer}}, \bibinfo {author}
  {\bibfnamefont {N.}~\bibnamefont {Camus}}, \bibinfo {author} {\bibfnamefont
  {K.~J.}\ \bibnamefont {Betsch}}, \bibinfo {author} {\bibfnamefont
  {O.}~\bibnamefont {Herrwerth}}, \bibinfo {author} {\bibfnamefont
  {A.}~\bibnamefont {Senftleben}}, \bibinfo {author} {\bibfnamefont {A.~M.}\
  \bibnamefont {Sayler}}, \bibinfo {author} {\bibfnamefont {T.}~\bibnamefont
  {Rathje}},  \emph {et~al.},\ }\href@noop {} {\bibfield  {journal} {\bibinfo
  {journal} {Nature communications}\ }\textbf {\bibinfo {volume} {3}},\
  \bibinfo {pages} {813} (\bibinfo {year} {2012}{\natexlab{b}})}\BibitemShut
  {NoStop}%
\bibitem [{\citenamefont {Brabec}\ \emph {et~al.}(1996)\citenamefont {Brabec},
  \citenamefont {Ivanov},\ and\ \citenamefont {Corkum}}]{brabec_pra96}%
  \BibitemOpen
  \bibfield  {author} {\bibinfo {author} {\bibfnamefont {T.}~\bibnamefont
  {Brabec}}, \bibinfo {author} {\bibfnamefont {M.~Y.}\ \bibnamefont {Ivanov}},
  \ and\ \bibinfo {author} {\bibfnamefont {P.~B.}\ \bibnamefont {Corkum}},\
  }\href@noop {} {\bibfield  {journal} {\bibinfo  {journal} {Physical Review
  A}\ }\textbf {\bibinfo {volume} {54}},\ \bibinfo {pages} {R2551} (\bibinfo
  {year} {1996})}\BibitemShut {NoStop}%
\bibitem [{\citenamefont {Comtois}\ \emph {et~al.}(2005)\citenamefont
  {Comtois}, \citenamefont {Zeidler}, \citenamefont {P{\'e}pin}, \citenamefont
  {Kieffer}, \citenamefont {Villeneuve},\ and\ \citenamefont
  {Corkum}}]{pepin_jpb05}%
  \BibitemOpen
  \bibfield  {author} {\bibinfo {author} {\bibfnamefont {D.}~\bibnamefont
  {Comtois}}, \bibinfo {author} {\bibfnamefont {D.}~\bibnamefont {Zeidler}},
  \bibinfo {author} {\bibfnamefont {H.}~\bibnamefont {P{\'e}pin}}, \bibinfo
  {author} {\bibfnamefont {J.}~\bibnamefont {Kieffer}}, \bibinfo {author}
  {\bibfnamefont {D.}~\bibnamefont {Villeneuve}}, \ and\ \bibinfo {author}
  {\bibfnamefont {P.}~\bibnamefont {Corkum}},\ }\href@noop {} {\bibfield
  {journal} {\bibinfo  {journal} {Journal of Physics B: Atomic, Molecular and
  Optical Physics}\ }\textbf {\bibinfo {volume} {38}},\ \bibinfo {pages} {1923}
  (\bibinfo {year} {2005})}\BibitemShut {NoStop}%
\bibitem [{\citenamefont {Arkhipov}\ \emph {et~al.}(2023)\citenamefont
  {Arkhipov}, \citenamefont {Arkhipov}, \citenamefont {Pakhomov}, \citenamefont
  {Obraztsov},\ and\ \citenamefont {Rosanov}}]{arkhipov_jetpl23}%
  \BibitemOpen
  \bibfield  {author} {\bibinfo {author} {\bibfnamefont {R.~M.}\ \bibnamefont
  {Arkhipov}}, \bibinfo {author} {\bibfnamefont {M.~V.}\ \bibnamefont
  {Arkhipov}}, \bibinfo {author} {\bibfnamefont {A.~V.}\ \bibnamefont
  {Pakhomov}}, \bibinfo {author} {\bibfnamefont {P.~A.}\ \bibnamefont
  {Obraztsov}}, \ and\ \bibinfo {author} {\bibfnamefont {N.~N.}\ \bibnamefont
  {Rosanov}},\ }\href@noop {} {\bibfield  {journal} {\bibinfo  {journal} {JETP
  Letters}\ }\textbf {\bibinfo {volume} {117}},\ \bibinfo {pages} {8} (\bibinfo
  {year} {2023})}\BibitemShut {NoStop}%
\bibitem [{\citenamefont {Sainadh}\ \emph {et~al.}(2019)\citenamefont
  {Sainadh}, \citenamefont {Xu}, \citenamefont {Wang}, \citenamefont
  {Atia-Tul-Noor}, \citenamefont {Wallace}, \citenamefont {Douguet},
  \citenamefont {Bray}, \citenamefont {Ivanov}, \citenamefont {Bartschat},
  \citenamefont {Kheifets} \emph {et~al.}}]{litvibyuk_nat19}%
  \BibitemOpen
  \bibfield  {author} {\bibinfo {author} {\bibfnamefont {U.~S.}\ \bibnamefont
  {Sainadh}}, \bibinfo {author} {\bibfnamefont {H.}~\bibnamefont {Xu}},
  \bibinfo {author} {\bibfnamefont {X.}~\bibnamefont {Wang}}, \bibinfo {author}
  {\bibfnamefont {A.}~\bibnamefont {Atia-Tul-Noor}}, \bibinfo {author}
  {\bibfnamefont {W.~C.}\ \bibnamefont {Wallace}}, \bibinfo {author}
  {\bibfnamefont {N.}~\bibnamefont {Douguet}}, \bibinfo {author} {\bibfnamefont
  {A.}~\bibnamefont {Bray}}, \bibinfo {author} {\bibfnamefont {I.}~\bibnamefont
  {Ivanov}}, \bibinfo {author} {\bibfnamefont {K.}~\bibnamefont {Bartschat}},
  \bibinfo {author} {\bibfnamefont {A.}~\bibnamefont {Kheifets}},  \emph
  {et~al.},\ }\href@noop {} {\bibfield  {journal} {\bibinfo  {journal}
  {Nature}\ }\textbf {\bibinfo {volume} {568}},\ \bibinfo {pages} {75}
  (\bibinfo {year} {2019})}\BibitemShut {NoStop}%
\end{thebibliography}%
\end{document}